\documentclass[12pt,a4paper]{article}
\usepackage[utf8]{inputenc}
\usepackage[left=18mm,top=20mm,right=18mm,bottom=15mm]{geometry}
\usepackage[english]{babel}
\usepackage{mathrsfs}
\usepackage{amsmath}
\usepackage{amsfonts}
\usepackage{amssymb}
\usepackage{makeidx}
\usepackage{graphicx}
\usepackage{physics}
\usepackage{float} 
\usepackage{subfloat}
\usepackage{multicol}
\usepackage{subcaption}
\usepackage{color}
\usepackage[dvipsnames]{xcolor}
\usepackage{enumerate}
\usepackage[colorlinks]{hyperref}
\hypersetup{
	colorlinks,
	citecolor=red,
	linkcolor=red,
	urlcolor=violet}
\pagecolor[rgb]{1.0,0.99,1.0}
\usepackage{wrapfig}   
\usepackage{paralist}

\usepackage{authblk}

\author{Jitendra Pal\thanks{jpal1@ph.iitr.ac.in}}

\affil{Department of Physics, Indian Institute of Technology Roorkee, Roorkee
	247667 Uttarakhand, India}

\begin{document}
	\date{}
	\title{{\bf{Chaotic string dynamics in Bosonic $\eta$-deformed $AdS_5 \times T^{ 1,1}$} background}}
	\maketitle
	\begin{abstract}\label{abstract}
		We investigate a new class of $\eta$-deformed $AdS_5 \times T^{1,1}$ backgrounds produced by classical  $r$-matrices that satisfy the modified classical Yang-Baxter equation [Jour. High Ener. Phys.\text{03} (2022) 094]. We consider the string sigma model on $\eta$-deformed $AdS_5 \times T^{1,1}$ background. We numerically examined the classical phase space dynamics of these (semi)classical strings over this deformed background and computed several chaos indicators.  These involve figuring out the Poincaré section and computing the Lyapunov exponents  . In the (semi)classical limit, we discover evidence that supports a non-integrable phase space dynamics.
	\end{abstract}
	\newpage
	\section{Introduction and summary}\label{intro}

The phase space of the majority of the dynamical systems is not integrable in the context of integrability; as a result, the role of chaotic classical dynamics has been extensively studied in the past. One of the main areas in string theory is the AdS/CFT correspondence. In the Gauge/Gravity Correspondence context, chaos was first studied in \cite{AdS/CFT chaos}. The duality between type IIB string theory on $AdS_5\times S_5$ and the $N = 4 $ SU(N) super Yang-Mills theory in the large N limit is a typical illustration of the AdS/CFT correspondence\cite{maldacena, Witten, GKP}. The internal manifold $S_5$ can be swapped with an Einstein manifold $T^{ 1,1}$ that also maintains conformal symmetry. Contrary to $AdS_5 \times S^5$ and $AdS_4 \times CP^3$, the classical integrability of the superstring on $AdS_5 \times T^{1,1}$ is not as manifest as in these cases. It has been shown that the superstring on $AdS_5 \times T^{1,1}$ is not classically integrable\cite{PDA1,PDA2} in the same sense as $AdS_5 \times S^5$\cite{bena}. Also, the string motion in the most general Sasaki-Einstein spaces, the $L^{a,b,c}$ Sasaki-Einstein manifolds, are non-integrable and chaotic \cite{g1}. The $T^{1,1}$ \cite{coni1, coni2,coni3} manifold is just a special case of the most general $ L^{a,b,c}$ models for the particular values  $a=b=c=1$.

		Integrable dynamical systems are mathematical models that can be solved exactly using analytical methods. These systems are characterized by having a sufficient number of conserved quantities, also called integrals of motion, that allow their solutions to be expressed in terms of elementary functions. These conserved quantities arise from symmetries of the system, and their existence can often be linked to the presence of certain algebraic or geometric structures in the system. In a dynamical system, the behavior of a physical system is described by a set of differential equations. Integrable systems are those for which it is possible to find explicit solutions that satisfy the differential equations, allowing for a complete understanding of the system's behavior. However, using the idea of Kovacic's algorithm \cite{analitic1,analitic2,analitic3} based on a set of essential but insufficient conditions, the Liouvillian (non-)integrability criteria for a typical 2d sigma model over general backgrounds can be analytically checked.
		
		Because integrability is not a common quality, exploring the entire  variety of integrable models is difficult to accomplish. As a result, approaches for deforming integrable theories while retaining their integrability have been built. There has been significant advancement in the systematic discussion of 2D non-linear sigma model integrable deformations, which includes the Yang-Baxter deformation. These are based on the solutions of Yang-Baxter equation, it can be homogeneous classical Yang-Baxter equation
		 (hCYBE)\cite{hY1,hY2} or modified classical Yang-Baxter equation(mCYBE)\cite{mY1,mY2} and further generalised to the symmetric and semi-symmetric coset spaces \cite{scY1,scY2,scY3}. Specifically, the  Yang-Baxter deformation based on the mCYBE is frequently referred to as the $\eta$-deformation. In the context of integrability, the analytic non-integrability and chaotic behaviour of $\eta$-deformed  $AdS_5 \times S^5$ are discussed in \cite{analytic1, chaos1}.

		 Yang-Baxter deformations of $T ^{1,1}$ have been addressed in \cite{ D1,D2, D3}. Applying
the TsT transformation to $AdS_5 \times T^{1,1}$ gives rise to the so called $\gamma$-deformed  $AdS_5 \times T^{1,1}$ discussed in \cite{ D4, D5}. A deformation of the traditional classical Yang-Baxter sigma model having classical r-matrices satisfying
the classical Yang-Baxter equation (CYBE), as described in \cite{D1}, has also been used to generate the above  $\gamma$-deformed geometry. The chaotic behaviour of this background is studied in \cite{pani}. Also, ABL model a fascinating generalization of the  $T ^{1,1}$ geometry found in \cite{Arutyunov} is integrable with NSNS flux containing a critical value of
nonvanishing  B-field. The numerical and analytic integrability of ABL model has been checked in \cite{C1, C2}.

 Bosonic $\eta$-deformed $AdS_5 \times T^{1,1}$\cite{eta} are generated by those classical r-matrix that satiesfies the modified Yang-Baxter equation.

\begin{align}\label{r-matrix}
	\begin{split}
		\left[ R(X),R(Y) \right] - R \qty(\left[R(X),Y \right]+\left[X,R(Y)\right])
		= c \left[X,Y \right]  \, ,
	\end{split}
\end{align}

where R is a constant linear constant on $\mathfrak{g}$ is connected to the classical $r$-matrix, where   $X, Y \in \mathfrak{g}  $  and $ \mathfrak{g}$ is the Lie algebra associated with Lie group G. For the mCYBE,
$c=\pm 1$ whereas, for CYBE $c=0$.

The study of chaotic dynamics in deformed string theory backgrounds is an active area of research, with many open questions and challenges. However, our example, the bosonic $\eta$-deformed $AdS_5 \times T^{1,1}$ provides a concrete example of a chaotic system in string theory, and sheds light on the behavior of strongly-coupled quantum systems in general.

The rest of the paper is structured as follows, deformed  $AdS_5\times T^{1,1}$ background geometry, fields and consistency of winding string embedding is disscussed in section2. The study of chaos in classical string dynamics is covered in section 3 using two separate approaches, first by examining the Poincaré section and secondly by computing the Lyapunov exponent. In section 4, we wrap things up with a few comments. Some large expressions that are included in the article's main text are collected in the  additional Appendices A.

\section{String sigma model on Bosonic $\eta$-deformed $AdS_5 \times T^{ 1,1}$ }
Let us begin by describing the geometry and general setup needed for our study. We start by writing the Bosonic $\eta$-deformed $AdS_5 \times T^{ 1,1}$ background \cite{eta}. It is sufficient to identify chaotic motions to show non-integrability. For that we consider string sigma model with the consistent string embedding. 

\subsection{$\eta$-deformed $AdS_5 \times T^{ 1,1}$ background }

The $\eta$-deformed $A d S_5$  metric is

\begin{align}\label{AdS}
	\frac{d s_{A d S_5}^2}{\left(1+\chi^2\right)}=-\frac{\left(1+\rho^2\right) d t^2}{1+\chi^2 q_1^2}+\frac{d \rho^2}{\left(1+\rho^2\right)\left(1+\chi^2 q_1^2\right)}+\rho^2\left(\frac{d \zeta^2+\cos ^2 \zeta d \psi_1^2}{1+\chi^2 q_2^2}+\sin ^2 \zeta d \psi_2^2\right),
\end{align}
with $\eta$-deformed $T^{1,1}$ metric is
\begin{align}\label{T11}
	\frac{d s_{T^{1,1}}^2}{\left(1+\chi^2\right)}=\mathcal{F}_3( & \frac{1}{6}\left(G\left(q_4^2+q_6^2 \mid 0\right) d \theta_1^2+G\left(q_3^2+q_5^2 \mid 0\right) d \theta_2^2\right. \nonumber\\
	& \left.+G\left(q_4^2+q_5^2+q_6^2 \mid q_4^2 q_5^2\right) \sin ^2 \theta_1 d \phi_1^2+G\left(q_3^2+q_5^2+q_6^2 \mid q_3^2 q_6^2\right) \sin ^2 \theta_2 d \phi_2^2\right) \nonumber\\
	& \left.+\frac{1}{9} G\left(q_3^2 \mid 0\right) G\left(q_4^2 \mid 0\right)\left(\cos \theta_1 d \phi_1+\cos \theta_2 d \phi_2+d \psi\right)^2\right),
\end{align}
where,

\begin{align*}
    \mathcal{F}_3^{-1}=G\big(q_3^2+q_4^2+q_5^2+q_6^2 \mid q_3^2 q_4^2+q_3^2 q_6^2+q_4^2 q_5^2\big).
\end{align*}

Each sector additionally has a $B$-field because of the $\eta$-deformation.
\begin{align}
	\frac{B_{A d S_5}}{\left(1+\chi^2\right)}=\frac{2 i \chi q_1^2}{3\left(1+\chi^2 q_1\right)} d t \wedge d \rho-\frac{2 \chi q_2 \rho \cos \zeta}{3\left(1+\chi^2 q_2^2\right)} d \zeta \wedge d \psi_1,
\end{align}

and 
\begin{align}\label{B}
	\frac{B_{T^{1,1}}}{\left(1+\chi^2\right)}=-\frac{\chi \mathcal{F}_3}{9}( & \left(\sqrt{6} q_5 G\left(q_4^2 \mid 0\right) \cos \theta_1+3 q_3 G\left(q_4^2+q_6^2 \mid 0\right) \sin \theta_1\right) d \theta_1 \wedge d \phi_1\nonumber \\
	& +q_5\left(\sqrt{6} G\left(q_4^2 \mid 0\right) \cos \theta_2+3 q_4 q_6 \chi^2 \sin \theta_2\right) d \theta_1 \wedge d \phi_2\nonumber \\
	& +q_6\left(\sqrt{6} G\left(q_5^2 \mid 0\right) \cos \theta_1-3 q_3 q_5 \chi^2 \sin \theta_1\right) d \theta_2 \wedge d \phi_1\nonumber \\
	& +\left(\sqrt{6} q_6 G\left(q_3^2 \mid 0\right) \cos \theta_2-3 q_4 G\left(q_3^2+q_5^2 \mid 0\right) \sin \theta_2\right) d \theta_2 \wedge d \phi_2\nonumber \\
	& \left.+\sqrt{6} q_5 G\left(q_4^2 \mid 0\right) d \theta_1 \wedge d \psi+\sqrt{6} q_6 G\left(q_3^2 \mid 0\right) d \theta_2 \wedge d \psi\right)
\end{align}
 
 and
 
 \begin{align}
 	G(r \mid s)=1+r \chi^2+s\chi^4
 \end{align}
 
where

 $q_1=i\rho,\ q_2=-\rho^2\sin\xi,\   q_3=-\cos\theta_1,\ q_4=\cos\theta_2,\ q_5=\sqrt{\frac{2}{3}}\sin\theta_1,\ q_6=\sqrt{\frac{2}{3}}\sin\theta_2$.

\subsection{Basic setup}
With this deformed background, we use the Polyakov action coupled with an antisymmetric B-field to study the chaotic dynamics. The $2d$ worldsheet string sigma model in the conformal gauge can be written as 
\begin{equation}
	\label{POL}
	S_{P}=-\frac{1}{2}\int \dd\tau\dd\sigma \left( \eta^{ab}G_{MN}+\epsilon^{ab}
	B_{MN}\right)\partial_{a}X^{M}\partial_{b}X^{N} \;=\int \dd\tau\dd\sigma\mathcal{L}_p ,
\end{equation}

where, $\displaystyle \eta_{ab}=\text{diag}\left(-1,1\right)$ indicate the world-sheet metric with coordinates $(\tau,\sigma)$, $G_{MN}$ and $B_{MN}$ denote the background metric and B-field, and $X_M$ be the target-space coordinates, where,  $(M, N= t, \theta_1, \theta_2,\phi_1,\phi_2,\psi)$. Here, we consider only $t$ coordinate of $AdS_5$ and all other coordinates are of $ T^{ 1,1}$  $( \theta_1, \theta_2,\phi_1,\phi_2,\psi)$ and for Levi-Civita symbol, we take the following convention: $\epsilon^{\tau\sigma}=-1$.

From the Polyakov action (\ref{POL}), the conjugate momenta related to the target space coordinates $X_M$ can be calculated as

\begin{equation}\label{mom}
	p_{M}=\frac{\partial \mathcal{L}_{P}}{\partial\dot{X}^{M}}=
	G_{MN}\partial_{\tau}X^{N}+B_{MN}\partial_{\sigma}X^{N} \, .
\end{equation}

Also, we can obtain the general expression of energy momentum stress tensor from Polyakov action (\ref{POL})
\begin{equation}\label{st}
	T_{ab} = \frac{1}{2}\qty(G_{MN}\partial_{a}X^{M}\partial_{b}X^{N}
	-\frac{1}{2}h_{ab}h^{cd}G_{MN}\partial_{c}X^{M}\partial_{d}X^{N}) \, ,
\end{equation}
where $h_{ab}=e^{2\omega\qty(\tau,\sigma)}\eta_{ab}$ in the conformal gauge. Also, The Virasoro constraints satisfies
\begin{align}\label{Vir:EM}
	\begin{split}
		T_{\tau\tau} = T_{\sigma\sigma} &=0 \, ,  \\
		T_{\tau\sigma} = T_{\sigma\tau} &=0 \, .
	\end{split}
\end{align}

Further, $T_{\tau\tau} $ component of stress tensor (\ref{st}) precisely gives Hamiltonian. The expression for the Hamiltonian of the system in terms of target space coordinates $X_M$ can be written as
\begin{equation}\label{Hamil}
	\mathcal{H} = p_{M}\partial_{\tau}X^{M}-\mathcal{L}_{P}=
	\frac{1}{2}G_{MN}\qty(\partial_{\tau}X^{M}\partial_{\tau}X^{N}
	+\partial_{\sigma}X^{M}\partial_{\sigma}X^{N}) \, .
\end{equation}

As seen in Polyakov action (\ref{POL}), the Lagrangian density can be expressed as
\begin{align}
	\label{lag}
	\mathcal{L}_p&=-\frac{\mathcal{D}_1}{12} \big(\chi ^2+1\big)\bigg(  6\big(\dot{t}^2 -t'^2\big) \big(4 \chi ^2 \sin ^2\theta _1+4 \chi ^2
	\sin ^2\theta _2+\chi ^2 \cos ^2\theta _1 \big(\chi ^2 \cos2 \theta _2\nonumber\\&+5 \chi ^2+6\big)+\cos ^2\theta _2
	\big(4 \chi ^4 \sin ^2\theta _1+6 \chi ^2\big)+6\big)+\big(\theta_1'^2-\dot{\theta}_1 ^2\big) \big(\chi ^2 \cos
	2 \theta _2+5 \chi ^2+6\big)\nonumber\\&+\big(\theta_2'^2-\dot{\theta}_2 ^2\big) \big(\chi ^2 \cos
	2 \theta _1+5 \chi ^2+6\big)
	+2 \big(\phi_1'^2-\dot{\phi}_1^2\big) \big(2 \cos ^2\theta _1 \big(\chi ^2 \cos\theta _2+1\big)\nonumber\\&+\sin ^2\theta _1
	\big(2 \chi ^2 \sin ^2\theta _1+2 \chi ^2 \sin ^2\theta _2+\chi ^2 \cos ^2\theta _2 \big(2 \chi ^2 \sin ^2\theta
	_1+3\big)+3\big)\nonumber\\&-2 \cos ^3\theta _1 \big(\chi ^4 \cos\theta _2+\chi ^2\big)\big)
	+2 \big(\phi_2'^2-\dot{\phi}_2^2\big) \big(-2 \chi ^2 \cos
	^3\theta _2 \big(\chi ^2 \cos \theta _1-1\big)\nonumber\\&+\cos ^2\theta _2 \big(2-2 \chi ^2 \cos \theta _1\big)+\sin
	^2\theta _2 \big(2 \chi ^2 \sin ^2\theta _1+2 \chi ^2 \sin ^2\theta _2+\chi ^2 \cos ^2\theta _1 \big(2 \chi ^2
	\sin ^2\theta _2\nonumber\\&+3\big)+3\big)\big)
	-4 \big(\psi '^2-\dot{\psi}^2\big)	\big(\chi ^2 \cos \theta _1-1\big) \big(\chi ^2 \cos\theta _2+1\big)	-8 \cos \theta _1 \cos\theta _2 \big(\chi ^2 \cos \theta _1\nonumber\\&-1\big)
	\big(\chi ^2 \cos\theta _2+1\big) \big(\phi _1' \phi _2'-\dot{\phi}_2 \dot{\phi}_1\big)
	-8 \cos \theta _1 \big(\chi ^2 \cos \theta _1-1\big) \big(\chi ^2 \cos\theta _2+1\big)
	\big(\psi ' \phi _1'\nonumber\\&-\dot{\psi} \dot{\phi}_1\big)
	-8 \cos\theta _2 \big(\chi ^2 \cos \theta _1-1\big) \big(\chi
	^2 \cos\theta _2+1\big) \big(\psi ' \phi _2'-\dot{\psi} \dot{\phi}_2\big)\bigg)
	\nonumber\\&	-\frac{\mathcal{D}_1}{18} \chi  \big(\chi ^2+1\big) \bigg(3\dot{\theta}_1 \big(8 \sin \theta _1 \big(\psi ' \left(\chi ^2 \cos \theta _2+1\right)+\cos
	\theta _2 \phi _2' \big(\chi ^2 \sin ^2\theta _2+\chi ^2 \cos \theta _2+1\big)\big)\nonumber\\&+\sin 2\theta _1 \phi
	_1' \big(4 \chi ^2 \cos \theta _2-\chi ^2 \cos 2\theta _2-5 \chi ^2-2\big)\big)-\dot{\theta}_2 \big(24 \sin \theta
	_2 \psi ' \big(\chi ^2 \cos \theta _1-1\big)\nonumber\\&+4 \sin \theta _2 \cos \theta _1 \phi _1' \big(-2 \sqrt{6} \chi ^2 \sin
	\theta _1+3 \chi ^2 \cos 2\theta _1-3 \chi ^2-6\big)+3 \sin 2\theta _2 \phi _2' \big(4 \chi ^2 \cos \theta
	_1\nonumber\\&+\chi ^2 \cos 2\theta _1+5 \chi ^2+2\big)\big)\bigg),
\end{align}

where

\begin{align*}
\mathcal{D}_1=\frac{1}{	4 \chi ^2 \sin ^2\theta _1+4 \chi ^2 \sin ^2\theta _2+\chi ^2 \cos ^2\theta _1 \left(\chi ^2 \cos 2 \theta
	_2+5 \chi ^2+6\right)+\cos ^2\theta _2 \left(4 \chi ^4 \sin ^2\theta _1+6 \chi ^2\right)+6}.
\end{align*}

\subsection{Consistent truncation of system using winding string embedding }
In order to study the chaotic dynamics, worldsheet coordinates have to be given a consistent worldsheet embedding.  Let us take the winding string embedding \cite{C1}  in order simplify the system to a set of  differential equations:

\begin{align}
	\label{em}
	t &=t(\tau ) \, , & \theta_1 &= \theta_1(\tau) \, , & \theta_2 &=\theta_2(\tau) \, ,   \\[5pt]\nonumber
	\phi_1 &=\phi_1(\tau)+\alpha_1\sigma \, ,  & \phi_2 &=\phi_2(\tau)+
	\alpha_2\sigma \, ,  &\psi &=\psi(\tau)+\alpha_3\sigma \ .
\end{align}

The momentum components corresponding to the isometries also called conserved charges can be obtained from Lagrangian density (\ref{lag}) by substituting the embedding (\ref{em}), we get

\begin{subequations}
	\begin{align}
		\label{p1}
		p_{\phi_1}&=-\frac{\mathcal{D}_1}{3}\big(\chi ^2+1\big) \bigg(\dot{\phi_1}\big(-2 \cos ^2\theta _1\big(\chi ^2 \cos
		\theta _2+1\big)-\sin ^2\theta _1\big(2 \chi ^2 \sin ^2\theta _1+2 \chi ^2
		\sin ^2\theta _2\nonumber\\&+\chi ^2 \cos ^2\theta _2 \big(2 \chi ^2 \sin ^2\big(\theta
		_1\big)+3\big)+3\big)+2 \cos ^3\theta _1\big(\chi ^4 \cos \theta _2+\chi
		^2\big)\big)+2 \cos \theta _1\big(\chi ^2 \cos \theta _1\nonumber\\&-1\big) \big(\chi ^2 \cos
		\theta _2+1\big) \big(\dot{\psi}+\cos \theta _2 \dot{\phi_2}\big)\bigg), \\
	\label{p2}
	p_{\phi_2}&=-\frac{\mathcal{D}_1}{3}	\big(\chi ^2+1\big) \bigg(2 \cos \theta_2 \dot{\psi} \big(\chi ^2 \cos \theta_1-1\big)
	\big(\chi ^2 \cos \theta_2+1\big)+2 \cos \theta_1 \cos \theta_2 \dot{\phi}_1 \big(\chi ^2 \cos \theta_1\nonumber\\&-1\big) \big(\chi ^2 \cos \big(\theta
	_2\big)+1\big)+\dot{\phi}_2 \big(2 \chi ^2 \cos ^3\theta_2 \big(\chi ^2 \cos \theta
	_1-1\big)+2 \cos ^2\theta_2 \big(\chi ^2 \cos \theta_1-1\big)\nonumber\\&-\sin
	^2\theta_2 \big(2 \chi ^2 \sin ^2\theta_1+2 \chi ^2 \sin ^2\theta_2+\chi ^2
	\cos ^2\theta_1 \big(2 \chi ^2 \sin ^2\theta_2+3\big)+3\big)\big)\bigg), \\
	\label{p3}
	p_{\psi}&=	-\frac{2\mathcal{D}_1}{3} \big(\chi ^2+1\big) \big(\chi ^2 \cos \theta_1-1\big) \big(\chi ^2 \cos \theta
	_2+1\big) \big(\dot{\psi}+\cos \theta_1 \dot{\phi}_1+\cos\theta_2
	\dot{\phi}_2\big) .
	\end{align}
\end{subequations}

The Hamiltonian is exactly represented by the $(\tau,\tau )$-component of the stress-energy tensor$(T_{\tau,\tau}=H)$, and the Virasoro constraints therefore gives H = 0 the Hamiltonian constraint. The other independent component $T_{\tau\sigma}$ can be recast by substituting (\ref{mom}) in (\ref{st}) in the conformal gauge as

\begin{align}
	\label{ts}
	T_{\tau\sigma}&=p_M\partial_\sigma X^M\nonumber,\\&
	=p_t t' +p_{\phi_1}\phi_1'+p_{\phi_2}\phi_2'+p_{\psi}\psi'+p_{\theta_1} \theta_1'+p_{\theta_2} \theta_2',
\end{align}

Now for the consistency of the embedding (\ref{em}), the second virassoro constrained must vanish. Now substituting embedding (\ref{em}) into second virassoro constraint(\ref{ts}), we get 
\begin{align}
		T_{\tau\sigma}&
=p_{\phi_1}\alpha_1+p_{\phi_2}\alpha_2+p_{\psi}\alpha_3=0,
\end{align}

This expression leads one to the conclusion that this requirement can be meet by simply setting  $ p_{\phi_1} = p_{\phi_2} = p_{\psi} = 0$, which leads to following condition 
\begin{align}
	\dot{\phi_1}=	\dot{\phi_2}=	\dot{\psi}=0,
\end{align}

Here, we need to refine our string embedding for the consistency requirements(second virassoro constraint $T_{\tau\sigma}=0$ ). Hence, we propose a  refined string embedding for our example as follows

\begin{align}
	\label{em1}
	t &=t(\tau ) \, , & \theta_1 &= \theta_1(\tau) \, , & \theta_2 &=\theta_2(\tau) \, ,   \\[5pt]\nonumber
	\phi_1 &=\alpha_1\sigma \, ,  & \phi_2 &=
	\alpha_2\sigma \, ,  &\psi &=\alpha_3\sigma \ ,
\end{align}
which clearly satiesfies $T_{\tau\sigma}=0$.

The equation of motion for $t$

\begin{align}
    (1+\chi^2)   \partial_\tau(\dot{t})=0,
\end{align}

gives $\dot{t}=J_t$, we say it as energy. Non-trivial equations for $\theta_1$ and $\theta_2$ describing the motion of string  are given by

\begin{subequations}
\begin{align}
&\tilde{\mathcal{B}}(\theta _1,\theta _2)-6 \big(\chi ^2 \big(\cos 2 \theta _2 \chi ^2+5 \chi ^2+6\big) \cos ^2\theta _1+4 \chi ^2
	\sin ^2\theta _1+4 \chi ^2 \sin ^2\theta _2+\cos ^2\theta _2 \big(4 \sin ^2\theta _1 \chi ^4\nonumber\\&+6 \chi
	^2\big)+6\big) \big(-2 \sin 2 \theta _2 \dot{\theta}_1 \dot{\theta}_2 \chi ^2-2 \big(4 \cos \theta _1
	\big(\chi ^2 \alpha _2 \cos ^2\theta _2+\big(\sin ^2\theta _2 \alpha _2 \chi ^2+\alpha _3 \chi ^2+\alpha _2\big) \cos
	\theta _2\nonumber\\&+\alpha _3\big)-\cos 2 \theta _1 \big(-4 \cos \theta _2 \chi ^2+\cos 2 \theta _2 \chi ^2+5
	\chi ^2+2\big) \alpha _1\big) \dot{\theta}_1 \chi +4 \sin \theta _1 \sin \theta _2 \big(4 \cos \theta
	_1 \sin ^2\big(\frac{\theta _2}{2}\big) \alpha _1 \chi ^2\nonumber\\&+4 \cos \theta _2 \alpha _2 \chi ^2-3 \cos 2 \theta _2 \alpha
	_2 \chi ^2-\alpha _2 \chi ^2+2 \alpha _3 \chi ^2+2 \alpha _2\big) \dot{\theta}_2 \chi +\big(\cos 2 \theta _2 \chi ^2+5 \chi
	^2+6\big) \ddot{\theta}_1\big)=0,
\end{align}

\begin{align}
&\tilde{	\mathcal{C}}(\theta _1,\theta _2)	-\big(\chi ^2\big(\cos 2\theta _2 \chi ^2+5 \chi ^2+6\big)\cos
	^2\theta _1+4 \chi ^2 \sin ^2\theta _1+4 \chi ^2 \sin ^2\theta _2+\cos ^2\theta _2\big(4 \sin
	^2\theta _1 \chi ^4\nonumber\\&+6 \chi ^2\big)+6\big)\big(4 \chi \big(3\big(4 \cos \theta _1 \chi ^2+\cos 2\theta _1
	\chi ^2+5 \chi ^2+2\big)\cos 2\theta _2 \alpha _2+\cos \theta _2\big(3 \cos3 \theta _1\alpha _1 \chi ^2\nonumber\\&-2
	\sqrt{6} \sin 2\theta _1 \alpha _1 \chi ^2-12 \alpha _3-3 \cos \theta _1\big(\big(\chi ^2+4\big)\alpha _1-4 \chi ^2 \alpha
	_3\big)\big)\big)\dot{\theta}_2-4 \chi  \dot{\theta}_1\big(\sin \theta _2\big(4 \sqrt{6} \cos2 \theta
	_1\alpha _1 \chi ^2\nonumber\\&+6 \sin \theta _1\big(3 \cos ^2\theta _1 \alpha _1 \chi ^2-3 \sin ^2\theta _1 \alpha _1
	\chi ^2+\alpha _1 \chi ^2+2 \cos \theta _1 \cos \theta _2 \alpha _2 \chi ^2+2 \cos \theta _2 \alpha _2 \chi ^2\nonumber\\&+2
	\alpha _3 \chi ^2-2 \alpha _1\big)\big)+3 \chi  \sin 2\theta _1 \dot{\theta}_2\big)+6\big(\cos 2\theta _1
	\chi ^2+5 \chi ^2+6\big)\ddot{\theta}_2\big)=0.
\end{align}
\end{subequations}

\begin{figure}[H]
	\centering
	\hfill
	\begin{subfigure}{0.4\textwidth}
		\includegraphics[width=\textwidth]{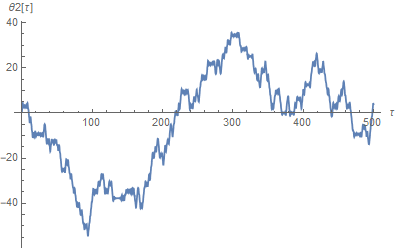}
		\caption{Energy=1, $\chi=1$}
		\label{fig:1}
	\end{subfigure}
	\hfill
	\begin{subfigure}{0.4\textwidth}
		\includegraphics[width=\textwidth]{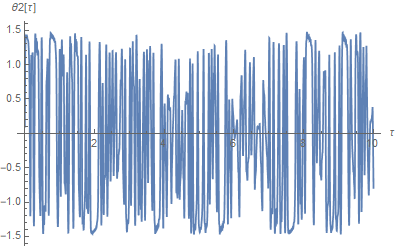}
		\caption{Energy=1, $\chi=20$}
		\label{fig:2}
	\end{subfigure}
	\caption{Plot of Solution for $\theta_2(\tau)$. The time evolution shows chaotic motion.}. 
\end{figure}

\section{Chaotic dynamics of string in deformed background}
In a Hamiltonian system with N degrees of freedom, the phase space is N-dimensional, and it is described by a set of coordinates $q_i$ and their conjugate momenta  $p_i$, where $ i=1,2,...,N$. The set$ (q_i,p_i)$ represents the state of the system at a given instant of time. If the system is integrable, then there are exactly N independent integrals of motion, which are conserved quantities that do not change as the system evolves in time. These integrals of motion can be expressed in terms of the coordinates $q_i$ and momenta $p_i$.
The KAM theorem is a fundamental result in the theory of dynamical systems that provides conditions under which a Hamiltonian system with an almost integrable Hamiltonian will exhibit a set of invariant tori in its phase space. These tori are important because they represent a subset of phase space where the motion is quasi-periodic and predictable, bridging the gap between integrable and chaotic systems. When the KAM torus is stable, the motion of the system on the torus can be represented by a periodic orbit in the Poincaré section. The periodic orbit corresponds to a closed curve in the section that intersects the section at regular intervals. As the perturbation becomes stronger, the periodic orbit can become unstable and break down, leading to the emergence of chaotic behavior. In the Poincaré section, this transition is characterized by the disappearance of the periodic orbit and the emergence of a chaotic set of Poincaré points.\\

In our scenario, we shall use the following approach. First, using the profile described in (\ref{em1}), we investigate the string motion on $AdS_5 \times T^{ 1,1}$ scenario. Beginning with the Lagrangian  $(\ref{lag})$ and string embedding mentioned in (\ref{em1}), we  write down the Hamiltonian as

\begin{align}
	\label{ham}
H&=	\frac{\mathcal{D}_1}{12}\big(\chi ^2+1\big) \bigg(-6 J_t^2 \big(4 \chi ^2 \sin ^2\theta_1+4 \chi ^2 \sin ^2\theta_2+\chi ^2 \cos ^2\theta_1 \big(\chi ^2 \cos 2\theta_2+5 \chi ^2+6\big)\nonumber\\&+\cos
	^2\theta_2 \big(4 \chi ^4 \sin ^2\theta_1+6 \chi ^2\big)+6\big)-4 \alpha _3^2
	\big(\chi ^2 \cos \theta_1-1\big) \big(\chi ^2 \cos \theta_2+1\big)\nonumber\\&-8 \alpha _1
	\alpha _2 \cos \theta_1 \cos \theta_2 \big(\chi ^2 \cos \theta_1-1\big)
	\big(\chi ^2 \cos \theta_2+1\big)-8 \alpha _1 \alpha _3 \cos \theta_1 \big(\chi ^2 \cos
	\theta_1-1\big) \big(\chi ^2 \cos \theta_2+1\big)\nonumber\\&-8 \alpha _2 \alpha _3 \cos
	\theta_2 \big(\chi ^2 \cos \theta_1-1\big) \big(\chi ^2 \cos \theta
	_2+1\big)+2 \alpha _2^2 \big(-2 \chi ^2 \cos ^3\theta_2 \big(\chi ^2 \cos \theta
	_1-1\big)\nonumber\\&+\cos ^2\theta_2 \big(2-2 \chi ^2 \cos \theta_1\big)+\sin
	^2\theta_2 \big(2 \chi ^2 \sin ^2\theta_1+2 \chi ^2 \sin ^2\theta_2+\chi ^2
	\cos ^2\theta_1 \big(2 \chi ^2 \sin ^2\theta_2+3\big)+3\big)\big)\nonumber\\&+2 \alpha _1^2
	\big(2 \cos ^2\theta_1 \big(\chi ^2 \cos \theta_2+1\big)+\sin ^2\theta_1
	\big(2 \chi ^2 \sin ^2\theta_1+2 \chi ^2 \sin ^2\theta_2+\chi ^2 \cos ^2\left(\theta
	_2\right) \big(2 \chi ^2 \sin ^2\theta_1+3\big)\nonumber\\&+3\big)-2 \cos ^3\theta_1 \big(\chi ^4
	\cos \theta_2+\chi ^2\big)\big)+\dot{\theta_1}^2 \big(\chi ^2 \cos \left(2 \theta
	_2\right)+5 \chi ^2+6\big)+\dot{\theta_2}^2 \big(\chi ^2 \cos\theta _1+5 \chi
	^2+6\big)\bigg).
\end{align}

The remaining components $p_{\theta_1}$ and $p_{\theta_2} $ can be obtained from plugging (\ref{em1}) in to (\ref{mom}), we get
\begin{subequations}
	\begin{align}
		\label{pt1}
		p_{\theta_1}&=\frac{\mathcal{D}_1}{6}	\big(\chi ^2+1\big) \bigg(-8 \chi  \sin \theta _1 \big(\alpha _3+\alpha _2 \chi ^2 \cos ^2\left(\theta
		_2\right)+\cos \theta _2 \left(\alpha _3 \chi ^2+\alpha _2+\alpha _2 \chi ^2 \sin ^2\theta
		_2\right)\big)\nonumber\\&+\alpha _1 \chi  \sin 2 \theta _1 \big(-4 \chi ^2 \cos \theta _2+\chi
		^2 \cos 2 \theta _2+5 \chi ^2+2\big)+\dot{\theta_1} \left(\chi ^2 \cos 2 \theta
		_2+5 \chi ^2+6\right)\bigg),
	\end{align}

	\begin{align}
		\label{pt2}
		p_{\theta_2}&=\frac{\mathcal{D}_1}{18}\big(\chi ^2+1\big) \bigg(3 \alpha _2 \chi  \sin 2\theta _2 \big(4 \chi ^2 \cos \theta _1+\chi ^2 \cos 2\theta _1+5 \chi ^2+2\big)+\chi  \sin \theta _2 \big(-4 \big(6
		\alpha _3\nonumber\\&+\sqrt{6} \alpha _1 \chi ^2 \sin 2 \theta _1\big)+6 \alpha _1 \chi ^2 \cos \left(3 \theta
		_1\right)-6 \big(\alpha _1 \left(\chi ^2+4\right)-4 \alpha _3 \chi ^2\big) \cos \theta _1\big)\nonumber\\&+3
		\dot{\theta_2} \big(\chi ^2 \cos 2\theta _1+5 \chi ^2+6\big)\bigg).
	\end{align}
\end{subequations}

We want to study the phase space dynamics with phase space coordinates  i.e.$(\theta_1,p_{\theta_1})$ and $(\theta_2,p_{\theta_2})$. Hence, now our next step is to compute the Hamilton’s equations. To get the Hamilton’s equations,  we substitute the expressions of $\dot{\theta_1}$ and $\dot{\theta_2}$ from (\ref{pt1},\ref{pt2}) into (\ref{ham}) to get the hamiltonian into  phase space coordinates$(\theta_1,p_{\theta_1},\theta_2,p_{\theta_2})$.
The resulting Hamilton’s equations of motion can be written as

\begin{subequations}
\begin{align}\label{theta1}
\dot{\theta_1}&=	\mathcal{D}_2 (\chi ^2+1)\Bigg[3p_{\theta_1} \bigg(\chi ^2 \big(\left(3 \chi ^2+2\right) \cos 2 \theta _2+\cos 2 \theta _1 \big(-\chi
^2 \cos 2 \theta _2+3 \chi ^2+2\big)\big)+\big(\chi ^2+2\big) \big(7 \chi
^2\nonumber\\&+6\big)\bigg)+ (\chi^3+\chi) \bigg(2 \sin  \theta _1 \big(2 \alpha _2 \chi ^2+4 \alpha
_3+\alpha _2 \chi ^2 \big(-\big(\cos 3 \theta _2-2 \cos 2 \theta
_2\big)\big)+\big(\big(\alpha _2+4 \alpha _3\big) \chi ^2\nonumber\\&+4 \alpha _2\big) \cos \theta
_2\big)-\alpha _1 \sin 2 \theta _1 \big(\chi ^2 \big(\cos 2 \theta _2-4 \cos
 \theta _2\big)+5 \chi ^2+2\big)\bigg)\bigg],
\end{align}
\begin{align}
	\dot{p}_{\theta_1}&=\mathcal{F}_1(p_{\theta_1},p_{\theta_2},\theta_1,\theta_2),
\end{align}
\begin{align}\label{theta2}
	\dot{\theta_2}&=3\mathcal{D}_4 \Bigg[9 p_{\theta_2} \bigg(\chi ^2 \big(\left(3 \chi ^2+2\right) \cos 2 \theta _2+\cos 2 \theta _1 \big(-\chi
	^2 \cos 2 \theta _2+3 \chi ^2+2\big)\big)+\big(\chi ^2+2\big) \big(7 \chi
	^2+6\big)\bigg)\nonumber\\&+\left(\chi ^3+\chi \right) \bigg(4 \sin \theta _2 \big(6 \alpha _3+\cos \left(\theta
	_1\right) \big(3 \left(\alpha _1-2 \alpha _3\right) \chi ^2+6 \alpha _1+\alpha _1 \chi ^2 \big(2 \sqrt{6} \sin
	 \theta _1\nonumber\\&-3 \cos 2 \theta _1\big)\big)\big)-3 \alpha _2 \sin 2 \theta _2
\big(\chi ^2 \left(4 \cos  \theta _1+\cos 2 \theta _1\right)+5 \chi ^2+2\big)\bigg)\bigg](\chi^2+1),
\end{align}
\begin{align}
	\dot{p}_{\theta_2}&=\mathcal{F}_2(p_{\theta_1},p_{\theta_2},\theta_1,\theta_2).
\end{align}
\end{subequations}

where, $\mathcal{D}_2$, $\mathcal{D}_4$,  $\mathcal{F}_1(p_{\theta_1},p_{\theta_2},\theta_1,\theta_2)$ and  $\mathcal{F}_2(p_{\theta_1},p_{\theta_2},\theta_1,\theta_2)$ expressions are given in Appendix.

\subsection{Numerical Analysis}

When a system exhibits chaotic behavior, its trajectory in the phase space becomes highly sensitive to initial conditions. This means that even small differences in the initial conditions can result in vastly different trajectories. As a result, chaotic systems often exhibit a highly complex and irregular motion in the phase space. By performing a numerical analysis of  Poincaré sections and Lyapunov exponents provides a powerful tool for analyzing the qualitative behavior of dynamical systems. By constructing Poincaré sections and calculating Lyapunov exponents for a wide range of initial conditions, one can gain insight into the various types of motion exhibited by the system, including periodic, quasi-periodic, and chaotic motion.

\subsubsection{Poincaré sections}
Poincaré sections are constructed by choosing a hyperplane in the phase space and then analyzing the intersection of the system's trajectory with the hyperplane. Each time the trajectory crosses the hyperplane, an intersection point is recorded. The resulting set of intersection points forms the Poincaré section, which is a two-dimensional subset of the phase space. 

Here, we have a four dimensional phase space$(\theta_1,\theta_2,p_{\theta_1},p_{\theta_2})$. It will be reduced to one dimension less because of the hamiltonian constraint(virassoro constraint, $T_{\tau\tau}=H=0$). String trajectory in the phase space are determined by the initial conditions applied to the phase space coordinates. We start with setting $p_{\theta_1}=0$ and fixing the values of $\theta_1  $ in the range $\theta_1(0){\displaystyle \in }  [0, 1]$ and varying $\theta_2  $ in the range $\theta_2(0){\displaystyle \in }  [0, 1]$  while maintaining energy constant  $J_t=\text{constant} $ and for fix value of deformation parameter $\chi$ , we obtain the corresponding values of $p_{\theta_2}$ for each plot. 

\begin{figure}[H]
\centering
\hfill
  \begin{subfigure}{0.4\textwidth}
    \includegraphics[width=\textwidth]{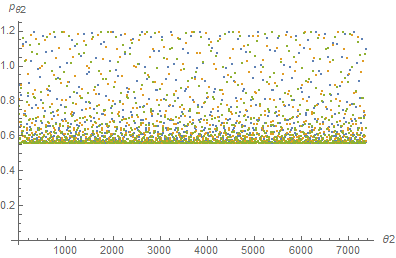}
    \caption{Energy=3}
    \label{fig:1}
  \end{subfigure}
 \hfill
  \begin{subfigure}{0.4\textwidth}
    \includegraphics[width=\textwidth]{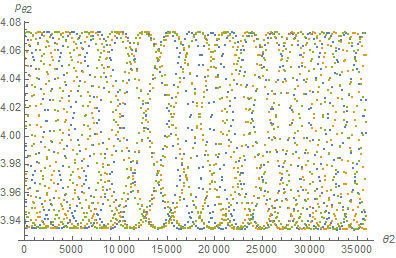}
    \caption{Energy=10}
    \label{fig:2}
  \end{subfigure}
   \caption{Poincaré sections for $\chi=0$}
\end{figure}

\begin{figure}[H]
\centering
\hfill
  \begin{subfigure}{0.4\textwidth}
    \includegraphics[width=\textwidth]{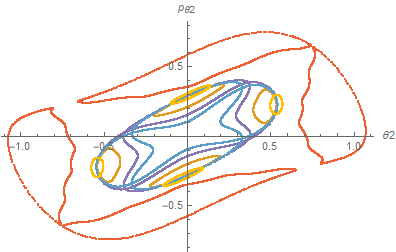}
    \caption{Energy=0.5}
    \label{fig:1}
  \end{subfigure}
 \hfill
  \begin{subfigure}{0.4\textwidth}
    \includegraphics[width=\textwidth]{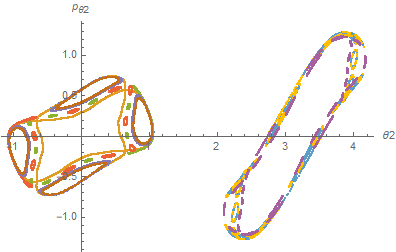}
    \caption{Energy=1}
    \label{fig:2}
  \end{subfigure}\\
   \hfill
  \begin{subfigure}{0.4\textwidth}
    \includegraphics[width=\textwidth]{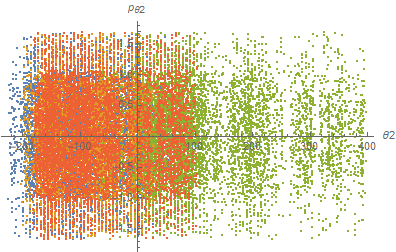}
    \caption{Energy=1.5}
    \label{fig:2}
  \end{subfigure}
   \hfill
  \begin{subfigure}{0.4\textwidth}
    \includegraphics[width=\textwidth]{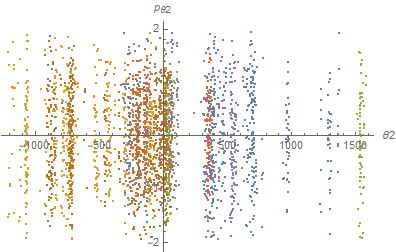}
    \caption{Energy=1.9}
    \label{fig:2}
  \end{subfigure}
   \caption{Poincaré sections for $\chi=1$}
\end{figure}

\begin{figure}[H]
\centering
  \begin{subfigure}{0.3\textwidth}
    \includegraphics[width=\textwidth]{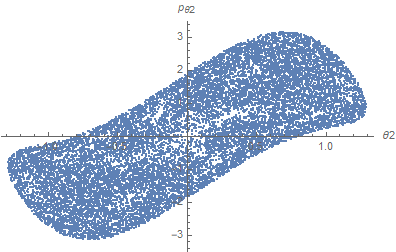}
    \caption{Energy=0.1}
    \label{fig:1}
  \end{subfigure}
  \begin{subfigure}{0.3\textwidth}
    \includegraphics[width=\textwidth]{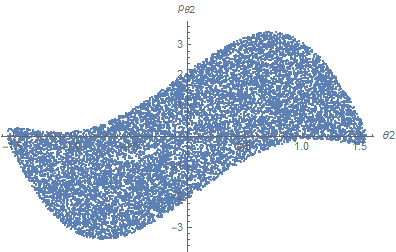}
    \caption{Energy=1}
    \label{fig:2}
  \end{subfigure}
  \begin{subfigure}{0.3\textwidth}
    \includegraphics[width=\textwidth]{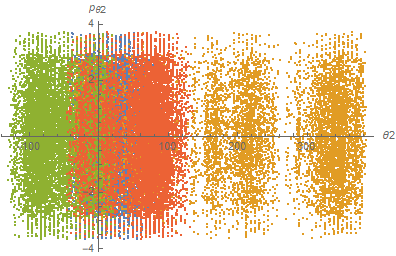}
    \caption{Energy=1.6}
    \label{fig:2}
  \end{subfigure}
   \caption{Poincaré sections for $\chi=2$}
\end{figure}

\begin{figure}[H]
\centering
  \begin{subfigure}{0.3\textwidth}
\includegraphics[width=\textwidth]{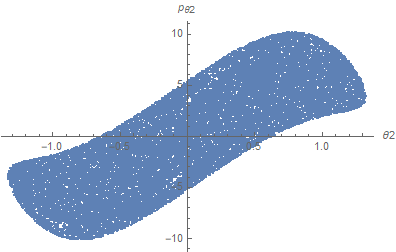}
    \caption{Energy=0.1}
    \label{fig:1}
  \end{subfigure}
  \begin{subfigure}{0.3\textwidth}
\includegraphics[width=\textwidth]{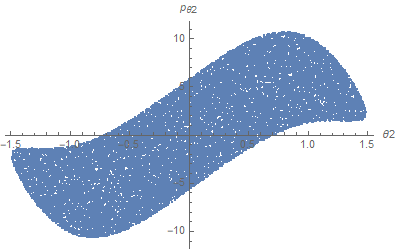}
    \caption{Energy=1}
    \label{fig:2}
  \end{subfigure}
  \begin{subfigure}{0.3\textwidth}
\includegraphics[width=\textwidth]{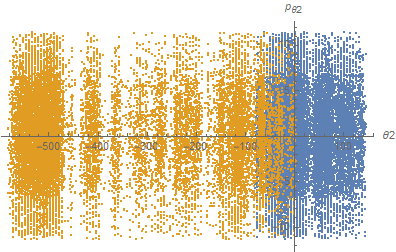}
    \caption{Energy=1.6}
    \label{fig:2}
  \end{subfigure}
   \caption{Poincaré sections for $\chi=5$}
\end{figure}

\begin{figure}[H]
\centering
  \begin{subfigure}{0.3\textwidth}
    \includegraphics[width=\textwidth]{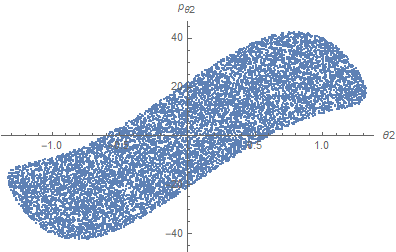}
    \caption{Energy=0.1}
    \label{fig:1}
  \end{subfigure}
  \begin{subfigure}{0.3\textwidth}
    \includegraphics[width=\textwidth]{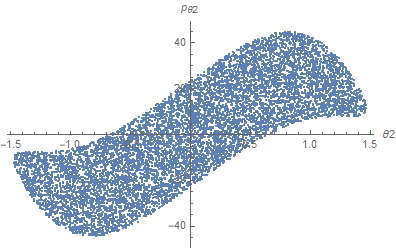}
    \caption{Energy=1}
    \label{fig:2}
  \end{subfigure}
  \begin{subfigure}{0.3\textwidth}
    \includegraphics[width=\textwidth]{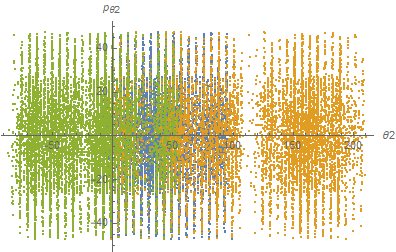}
    \caption{Energy=1.6}
    \label{fig:2}
  \end{subfigure}
   \caption{Poincaré sections for $\chi=20$}
\end{figure}

\begin{figure}[H]
\centering
  \begin{subfigure}{0.3\textwidth}
    \includegraphics[width=\textwidth]{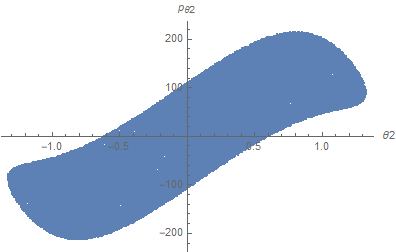}
    \caption{Energy=0.1}
    \label{fig:1}
  \end{subfigure}
  \begin{subfigure}{0.3\textwidth}
    \includegraphics[width=\textwidth]{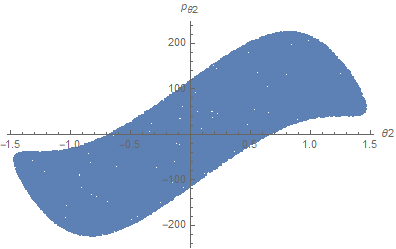}
    \caption{Energy=1}
    \label{fig:2}
  \end{subfigure}
  \begin{subfigure}{0.3\textwidth}
    \includegraphics[width=\textwidth]{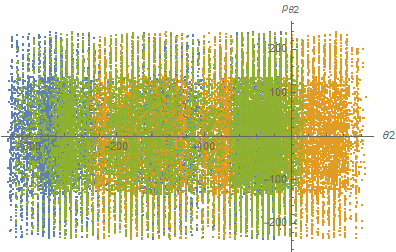}
    \caption{Energy=1.6}
    \label{fig:2}
  \end{subfigure}
   \caption{Poincaré sections for $\chi=100$}
\end{figure}

\subsection{ Lyapunov exponents}
Lyapunov exponents are mathematical quantities that describe the rate of separation of nearby trajectories in a dynamical system. They are used to measure the degree of chaos in a system and predict its long-term behavior. Let's take a look at two originally close-together orbits, one of which passes through the point$Y_0$ and the other $Y_0+\Delta Y_0$, where $\Delta Y (\tau)$ is the separation between two nearby trajectories at later time $t$ and $\left|\cdot\right|$ denotes the norm of a vector.  The expression for the Lyapunov exponents of a chaotic system is:

\begin{align}
    \lambda=\frac{1}{\tau} \ln \frac{\left\|\Delta Y\left(Y_0, \tau\right)\right\|}{\left\|\Delta Y_0\right\|},
\end{align}
The greatest Lyapunov exponent, which is measurable when the interval is quite large, should be taken.
\begin{align}
    \lambda_L=\lim _{\tau \rightarrow \infty} \frac{1}{\tau} \ln \frac{\left\|\Delta Y\left(Y_0, \tau\right)\right\|}{\left\|\Delta Y_0\right\|}=\lim _{\tau \rightarrow \infty} \frac{1}{\tau} \sum \lambda_i \tau_i.
\end{align}

 The Lyapunov exponents $\lambda_i$ are ordered such that $\lambda_1 \geq \lambda_2 \geq \cdots \geq \lambda_n$. If all the Lyapunov exponents are negative, then the system is stable. If there are one or more positive Lyapunov exponents, then the system is chaotic. It's value converges to a fixed positive number, indicating the system is chaotic as shown in Figure 8.

\begin{figure}[H]
\centering
\hfill
  \begin{subfigure}
  {0.4\textwidth}
    \includegraphics[width=\textwidth]{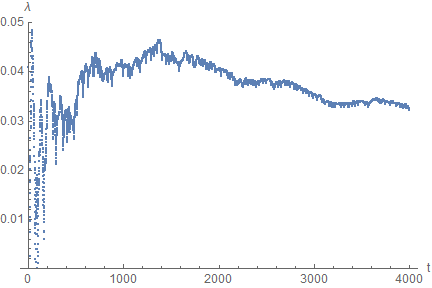}
    \caption{Energy=10, $\chi=0$}
    \label{fig:1}
  \end{subfigure}
 \hfill
  \begin{subfigure}
  {0.4\textwidth}
    \includegraphics[width=\textwidth]{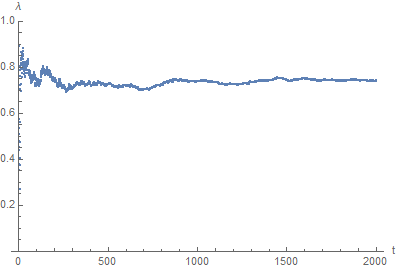}
    \caption{Energy=1.5, $\chi=1$}
    \label{fig:2}
  \end{subfigure}\\
   \hfill
  \begin{subfigure}{0.4\textwidth}
    \includegraphics[width=\textwidth]{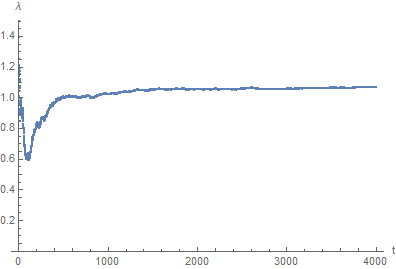}
    \caption{Energy=1, $\chi=2$}
    \label{fig:2}
  \end{subfigure}
   \hfill
  \begin{subfigure}{0.4\textwidth}
    \includegraphics[width=\textwidth]{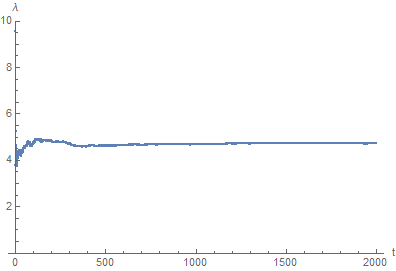}
    \caption{Energy=1, $\chi=5$}
    \label{fig:2}
  \end{subfigure}
   \caption{ Lyapunov exponents}
\end{figure}

\section{Conclusion}
In this article, we precisely shown that the string dynamics in the $\eta$-deformed $AdS_5 \times T^{ 1,1}$ geometry is non-integrable through the appearance of chaos. We examined the system's motion numerically and confirmed that its behaviour is chaotic by computing Poincare sections and Lyapunov exponents.
By looking at the Poincare sections for different values of deformation parameter $\chi$, the phase space trajectory is coming out to be scattered.
Also, the Lyapunov exponent converges to finite positive values in long time behaviour. 

One can check analytic integrability using Kovacic's algorithm.
Kovacic's algorithm is another way to determine whether the string sigma model is non-integrable over $\eta$-deformed $AdS_5 \times T^{ 1,1}$ background.

\section*{Acknowledgments}
The author is indebted to the authorities of IIT Roorkee
for their unconditional support towards researches in basic sciences. Author would
like to thank Arindam Lala,  Hemant Rathi and Manoranjan Samal for useful discussions.

\newpage
\appendix
\numberwithin{equation}{section}
\renewcommand{\theequation}{\thesection\arabic{equation}}
\section{ Some Detailed expressions }\label{Kova}

\begin{align*}
    \mathcal{D}_2=\frac{1}{(1+\chi^2)^2(6+5\chi^2+\chi^2\cos2\theta_2)}, \quad \mathcal{D}_3=\frac{1}{9(1+\chi^2)^2(6+5\chi^2+\chi^2\cos2\theta_1)^2},
\end{align*}

\begin{align*}
    \mathcal{D}_4=\frac{1}{(1+\chi^2)^2(6+5\chi^2+\chi^2\cos2\theta_1)},\quad \mathcal{D}_5=\frac{\mathcal{D}_1}{12},\quad  \mathcal{D}_6=\frac{1}{9(1+\chi^2)^2(6+5\chi^2+\chi^2\cos2\theta_2)^2}.
\end{align*}
\begin{align}
\tilde{\mathcal{B}}&=	6 \big(\chi  \big(-4 \cos \theta _2 \chi ^2+\cos 2 \theta _2 \chi ^2+5 \chi ^2+2\big) \sin 2 \theta _1 \alpha _1-8
	\chi  \sin \theta _1 \big(\chi ^2 \alpha _2 \cos ^2\theta _2+\big(\sin ^2\theta _2 \alpha _2 \chi ^2\nonumber\\&+\alpha _3 \chi
	^2+\alpha _2\big) \cos \theta _2+\alpha _3\big)+\big(\cos 2 \theta _2 \chi ^2+5 \chi ^2+6\big) \dot{\theta}_1\big) \big(\big(\cos 2 \theta _2 \chi ^2-3 \chi ^2-2\big) \sin 2 \theta _1 \dot{\theta}_1\nonumber\\&+\big(\cos \big(2
	\theta _1\big) \chi ^2-3 \chi ^2-2\big) \sin 2 \theta _2 \dot{\theta}_2\big) \chi ^2-\big(\cos 2 \theta _2 \chi
	^2-3 \chi ^2-2\big) \sin 2 \theta _1 \big(-18 \big(\chi ^2 \big(\cos 2 \theta _2 \chi ^2\nonumber\\&+5 \chi ^2+6\big) \cos
	^2\theta _1+4 \chi ^2 \sin ^2\theta _1+4 \chi ^2 \sin ^2\theta _2+\cos ^2\theta _2 \big(4 \sin
	^2\theta _1 \chi ^4+6 \chi ^2\big)+6\big) J_t^2\nonumber\\&-6 \big(-2 \big(\cos \theta _2 \chi ^4+\chi ^2\big) \cos ^3\theta
	_1+2 \big(\cos \theta _2 \chi ^2+1\big) \cos ^2\theta _1+\sin ^2\theta _1 \big(\chi ^2 \big(2 \chi ^2
	\sin ^2\theta _1\nonumber\\&+3\big) \cos ^2\theta _2+2 \chi ^2 \sin ^2\theta _1+2 \chi ^2 \sin ^2\theta
	_2+3\big)\big) \alpha _1^2-6 \big(-2 \chi ^2 \big(\chi ^2 \cos \theta _1-1\big) \cos ^3\theta _2+\big(2\nonumber\\&-2 \chi ^2
	\cos \theta _1\big) \cos ^2\theta _2+\sin ^2\theta _2 \big(\chi ^2 \big(2 \chi ^2 \sin ^2\theta
	_2+3\big) \cos ^2\theta _1+2 \chi ^2 \sin ^2\theta _1+2 \chi ^2 \sin ^2\theta _2\nonumber\\&+3\big)\big) \alpha
	_2^2+12 \big(\chi ^2 \cos \theta _1-1\big) \big(\cos \theta _2 \chi ^2+1\big) \alpha _3^2+3 \big(\cos 2 \theta
	_2 \chi ^2+5 \chi ^2+6\big) \dot{\theta}_1^2+3 \big(\cos 2 \theta _1 \chi ^2\nonumber\\&+5 \chi ^2+6\big) \dot{\theta}_2^2+24 \cos \theta _1 \big(\chi ^2 \cos \theta _1-1\big) \cos \theta _2 \big(\cos \theta _2 \chi
	^2+1\big) \alpha _1 \alpha _2+24 \cos \theta _1 \big(\chi ^2 \cos \theta _1\nonumber\\&-1\big) \big(\cos \theta _2 \chi
	^2+1\big) \alpha _1 \alpha _3+24 \big(\chi ^2 \cos \theta _1-1\big) \cos \theta _2 \big(\cos \theta _2 \chi
	^2+1\big) \alpha _2 \alpha _3-6 \chi  \big(8 \sin \theta _1 \big(\chi ^2 \alpha _2 \cos ^2\theta _2\nonumber\\&+\big(\sin ^2\theta
	_2\alpha _2 \chi ^2+\alpha _3 \chi ^2+\alpha _2\big) \cos \theta _2+\alpha _3\big)-\big(-4 \cos \theta _2 \chi
	^2+\cos 2 \theta _2 \chi ^2+5 \chi ^2+2\big) \sin 2 \theta _1 \alpha _1\big) \dot{\theta}_1\nonumber\\&+2 \chi  \big(4 \cos
	\theta _1 \big(3 \cos 2 \theta _1 \chi ^2-2 \sqrt{6} \sin \theta _1 \chi ^2-3 \big(\chi ^2+2\big)\big) \sin
	\theta _2 \alpha _1+3 \big(4 \cos \theta _1 \chi ^2+\cos 2 \theta _1 \chi ^2+5 \chi ^2\nonumber\\&+2\big) \sin 2 \theta
	_2 \alpha _2+24 \big(\chi ^2 \cos \theta _1-1\big) \sin \theta _2 \alpha _3\big) \dot{\theta}_2\big) \chi
	^2+2 \big(\chi ^2 \big(\cos 2 \theta _2 \chi ^2+5 \chi ^2+6\big) \cos ^2\theta _1\nonumber\\&+4 \chi ^2 \sin ^2\theta _1+4
	\chi ^2 \sin ^2\theta _2+\cos ^2\theta _2 \big(4 \sin ^2\theta _1 \chi ^4+6 \chi ^2\big)+6\big) \big(-9
	\big(\cos 2 \theta _2 \chi ^2-3 \chi ^2\nonumber\\&-2\big) \sin 2 \theta _1 J_t^2 \chi ^2-6 \sin \theta _1 \big(\chi ^2 \cos
	^3\theta _2+\cos ^2\theta _2-\cos \theta _1 \sin ^2\theta _2 \big(2 \chi ^2 \sin ^2\theta
	_2+1\big)\big) \alpha _2^2 \chi ^2\nonumber\\&-6 \big(\cos \theta _2 \chi ^2+1\big) \sin \theta _1 \alpha _3^2 \chi ^2-3 \sin
	2 \theta _1 \dot{\theta}_2^2 \chi ^2-6 \cos \theta _2 \big(\cos \theta _2 \chi ^2+1\big) \sin 2
	\theta _1 \alpha _1 \alpha _2 \chi ^2-6 \big(\cos \theta _2 \chi ^2\nonumber\\&+1\big) \sin 2 \theta _1 \alpha _1 \alpha _3 \chi
	^2-12 \cos \theta _2 \big(\cos \theta _2 \chi ^2+1\big) \sin \theta _1 \alpha _2 \alpha _3 \chi ^2-3 \big(8 \cos
	\theta _1 \big(\chi ^2 \alpha _2 \cos ^2\theta _2+\big(\sin ^2\theta _2 \alpha _2 \chi ^2\nonumber\\&+\alpha _3 \chi ^2+\alpha
	_2\big) \cos \theta _2+\alpha _3\big)-2 \cos 2 \theta _1 \big(-4 \cos \theta _2 \chi ^2+\cos 2 \theta
	_2 \chi ^2+5 \chi ^2+2\big) \alpha _1\big) \dot{\theta}_1 \chi \nonumber\\&-2 \sin \theta _2 \big(4 \sqrt{6} \cos 2 \theta
	_1 \alpha _1 \chi ^2+6 \sin \theta _1 \big(3 \cos ^2\theta _1 \alpha _1 \chi ^2-3 \sin ^2\theta _1 \alpha _1
	\chi ^2+\alpha _1 \chi ^2+2 \cos \theta _1 \cos \theta _2 \alpha _2 \chi ^2\nonumber\\&+2 \cos \theta _2 \alpha _2 \chi ^2+2
	\alpha _3 \chi ^2-2 \alpha _1\big)\big) \dot{\theta}_2 \chi -6 \cos \theta _1 \sin \theta _1 \big(4 \sin
	^2\theta _1 \chi ^2+2 \sin ^2\theta _2 \chi ^2-2 \cos \theta _2 \chi ^2\nonumber\\&+\cos ^2\theta _2 \big(4 \chi ^2
	\sin ^2\theta _1+3\big) \chi ^2+3 \cos \theta _1 \big(\cos \theta _2 \chi ^4+\chi ^2\big)+1\big) \alpha
	_1^2-12 \big(\chi ^2 \cos \theta _1-1\big) \cos \theta _2 \big(\cos \theta _2 \chi ^2\nonumber\\&+1\big) \sin \big(\theta
	_1\big) \alpha _1 \alpha _2-12 \big(\chi ^2 \cos \theta _1-1\big) \big(\cos \theta _2 \chi ^2+1\big) \sin \theta
	_1 \alpha _1 \alpha _3\big).
\end{align}

\begin{align}
\tilde{	\mathcal{C}}=&	\big(6 \chi \big(4 \cos \theta _1 \chi ^2+\cos 2\theta _1 \chi ^2+5 \chi ^2+2\big)\sin 2\theta _2 \alpha _2+4
	\chi  \sin \theta _2\big(3 \cos3 \theta _1\alpha _1 \chi ^2-2\big(\sqrt{6} \sin 2\theta _1 \alpha _1 \chi ^2+6
	\alpha _3\big)\nonumber\\&-3 \cos \theta _1\big(\big(\chi ^2+4\big)\alpha _1-4 \chi ^2 \alpha _3\big)\big)+6\big(\cos2 \theta
	_1\chi ^2+5 \chi ^2+6\big)\dot{\theta}_2\big)\big(\big(\cos 2\theta _2 \chi ^2-3 \chi ^2-2\big)\sin2
	\theta _1\dot{\theta}_1\nonumber\\&+\big(\cos 2\theta _1 \chi ^2-3 \chi ^2-2\big)\sin 2\theta _2 \dot{\theta}_2\big)\chi ^2-\big(\cos 2\theta _1 \chi ^2-3 \chi ^2-2\big)\sin 2\theta _2\big(-18\big(\chi ^2\big(\cos
	2\theta _2 \chi ^2\nonumber\\&+5 \chi ^2+6\big)\cos ^2\theta _1+4 \chi ^2 \sin ^2\theta _1+4 \chi ^2 \sin ^2\theta
	_2+\cos ^2\theta _2\big(4 \sin ^2\theta _1 \chi ^4+6 \chi ^2\big)+6\big)J_t^2\nonumber\\&-6\big(-2\big(\cos\theta
	_2\chi ^4+\chi ^2\big)\cos ^3\theta _1+2\big(\cos \theta _2 \chi ^2+1\big)\cos ^2\theta _1+\sin
	^2\theta _1\big(\chi ^2\big(2 \chi ^2 \sin ^2\theta _1+3\big)\cos ^2\theta _2\nonumber\\&+2 \chi ^2 \sin ^2\theta
	_1+2 \chi ^2 \sin ^2\theta _2+3\big)\big)\alpha _1^2-6\big(-2 \chi ^2\big(\chi ^2 \cos \theta _1-1\big)\cos
	^3\theta _2+\big(2-2 \chi ^2 \cos \theta _1\big)\cos ^2\theta _2\nonumber\\&+\sin ^2\theta _2\big(\chi ^2
	\big(2 \chi ^2 \sin ^2\theta _2+3\big)\cos ^2\theta _1+2 \chi ^2 \sin ^2\theta _1+2 \chi ^2 \sin ^2\theta
	_2+3\big)\big)\alpha _2^2+12\big(\chi ^2 \cos \theta _1-1\big)\big(\cos \theta _2 \chi ^2\nonumber\\&+1\big)\alpha _3^2+3
	\big(\cos 2\theta _2 \chi ^2+5 \chi ^2+6\big)\dot{\theta}_1^2+3\big(\cos 2\theta _1 \chi ^2+5 \chi ^2+6\big)
	\dot{\theta}_2^2+24 \cos \theta _1\big(\chi ^2 \cos \theta _1-1\big)\cos \theta _2\big(\cos
	\theta _2 \chi ^2\nonumber\\&+1\big)\alpha _1 \alpha _2+24 \cos \theta _1\big(\chi ^2 \cos \theta _1-1\big)\big(\cos
	\theta _2 \chi ^2+1\big)\alpha _1 \alpha _3+24\big(\chi ^2 \cos \theta _1-1\big)\cos \theta _2\big(\cos
	\theta _2 \chi ^2+1\big)\alpha _2 \alpha _3\nonumber\\&-6 \chi \big(8 \sin \theta _1\big(\chi ^2 \alpha _2 \cos ^2\theta
	_2+\big(\sin ^2\theta _2 \alpha _2 \chi ^2+\alpha _3 \chi ^2+\alpha _2\big)\cos \theta _2+\alpha _3\big)-\big(-4
	\cos \theta _2 \chi ^2+\cos 2\theta _2 \chi ^2\nonumber\\&+5 \chi ^2+2\big)\sin 2\theta _1 \alpha _1\big)\dot{\theta}_1+2 \chi \big(4 \cos \theta _1\big(3 \cos 2\theta _1 \chi ^2-2 \sqrt{6} \sin \theta _1 \chi ^2-3
	\big(\chi ^2+2\big)\big)\sin \theta _2 \alpha _1\nonumber\\&+3\big(4 \cos \theta _1 \chi ^2+\cos 2\theta _1 \chi ^2+5
	\chi ^2+2\big)\sin 2\theta _2 \alpha _2+24\big(\chi ^2 \cos \theta _1-1\big)\sin \theta _2 \alpha _3\big)
	\dot{\theta}_2\big)\chi ^2+2\big(\chi ^2\big(\cos 2\theta _2 \chi ^2\nonumber\\&+5 \chi ^2+6\big)\cos ^2\theta _1+4 \chi
	^2 \sin ^2\theta _1+4 \chi ^2 \sin ^2\theta _2+\cos ^2\theta _2\big(4 \sin ^2\theta _1 \chi ^4+6 \chi
	^2\big)+6\big)\big(-9\big(\cos 2\theta _1 \chi ^2-3 \chi ^2\nonumber\\&-2\big)\sin 2\theta _2 J_t^2 \chi ^2-6\big(\chi ^2 \cos
	^3\theta _1-\cos ^2\theta _1-\cos \theta _2 \sin ^2\theta _1\big(2 \chi ^2 \sin ^2\theta
	_1+1\big)\big)\sin \theta _2 \alpha _1^2 \chi ^2\nonumber\\&-6\big(\chi ^2 \cos \theta _1-1\big)\sin \theta _2
	\alpha _3^2 \chi ^2-3 \sin 2\theta _2 \dot{\theta}_1^2 \chi ^2-6 \cos \theta _1\big(\chi ^2 \cos\theta
	_1-1\big)\sin 2\theta _2 \alpha _1 \alpha _2 \chi ^2\nonumber\\&-12 \cos \theta _1\big(\chi ^2 \cos\theta
	_1-1\big)\sin \theta _2 \alpha _1 \alpha _3 \chi ^2-6\big(\chi ^2 \cos \theta _1-1\big)\sin2 \theta
	_2\alpha _2 \alpha _3 \chi ^2\nonumber\\&+12 \sin \theta _1 \sin \theta _2\big(4 \cos \theta _1 \sin
	^2\big(\frac{\theta _2}{2}\big)\alpha _1 \chi ^2+4 \cos \theta _2 \alpha _2 \chi ^2-3 \cos 2\theta _2 \alpha _2 \chi
	^2-\alpha _2 \chi ^2+2 \alpha _3 \chi ^2+2 \alpha _2\big)\dot{\theta}_1 \chi \nonumber\\&+\big(4 \cos \theta _1 \cos \theta _2
	\big(3 \cos 2\theta _1 \chi ^2-2 \sqrt{6} \sin \theta _1 \chi ^2-3\big(\chi ^2+2\big)\big)\alpha _1+6\big(4 \cos
	\theta _1 \chi ^2+\cos 2\theta _1 \chi ^2+5 \chi ^2\nonumber\\&+2\big)\cos 2\theta _2 \alpha _2+24\big(\chi ^2 \cos
	\theta _1-1\big)\cos \theta _2 \alpha _3\big)\dot{\theta}_2 \chi -6 \cos \theta _2 \sin\theta
	_2\big(2 \sin ^2\theta _1 \chi ^2+4 \sin ^2\theta _2 \chi ^2-3 \cos \theta _2 \chi ^2\nonumber\\&+\cos\theta
	_1\big(3 \cos \theta _2 \chi ^2+2\big)\chi ^2+\cos ^2\theta _1\big(4 \chi ^2 \sin ^2\theta _2+3\big)
	\chi ^2+1\big)\alpha _2^2-12 \cos \theta _1\big(\chi ^2 \cos \theta _1\nonumber\\&-1\big)\big(\cos \theta _2 \chi
	^2+1\big)\sin \theta _2 \alpha _1 \alpha _2-12\big(\chi ^2 \cos \theta _1-1\big)\big(\cos \theta _2 \chi
	^2+1\big)\sin \theta _2 \alpha _2 \alpha _3\big).
\end{align}

\begin{align}
\mathcal{F}_1&=	\big(\chi ^2+1\big) D_5 \big(\big(\cos 2\theta_2 \chi ^2-3 \chi ^2-2\big) \sin 2\theta_1 \big(-6 \big(\chi ^2
	\big(\cos 2\theta_2 \chi ^2+5 \chi ^2+6\big) \cos ^2\theta _1+4 \chi ^2 \sin ^2\theta _1\nonumber\\&+4 \chi ^2 \sin
	^2\theta _2+\cos ^2\theta _2 \big(4 \sin ^2\theta _1 \chi ^4+6 \chi ^2\big)+6\big) J_t^2+2 \big(-2 \big(\cos
	\theta _2 \chi ^4+\chi ^2\big) \cos ^3\theta _1+2 \big(\cos \theta _2 \chi ^2\nonumber\\&+1\big) \cos ^2\theta
	_1+\sin ^2\theta _1 \big(\chi ^2 \big(2 \chi ^2 \sin ^2\theta _1+3\big) \cos ^2\theta _2+2 \chi ^2 \sin
	^2\theta _1+2 \chi ^2 \sin ^2\theta _2+3\big)\big) \alpha _1^2\nonumber\\&+2 \big(-2 \chi ^2 \big(\chi ^2 \cos \theta
	_1-1\big) \cos ^3\theta _2+\big(2-2 \chi ^2 \cos \theta _1\big) \cos ^2\theta _2+\sin ^2\theta
	_2 \big(\chi ^2 \big(2 \chi ^2 \sin ^2\theta _2+3\big) \cos ^2\theta _1\nonumber\\&+2 \chi ^2 \sin ^2\theta _1+2 \chi
	^2 \sin ^2\theta _2+3\big)\big) \alpha _2^2-4 \big(\chi ^2 \cos \theta _1-1\big) \big(\cos \theta _2 \chi
	^2+1\big) \alpha _3^2+D_3 \big(9 \big(\big(\big(3 \chi ^2\nonumber\\&+2\big) \cos 2\theta_2+\cos 2\theta_1 \big(-\cos 2
	\theta _2 \chi ^2+3 \chi ^2+2\big)\big) \chi ^2+\big(\chi ^2+2\big) \big(7 \chi ^2+6\big)\big) p_{\theta_2}+2 \big(\chi ^3\nonumber\\&+\chi \big) \sin
	\theta _2 \big(-3 \big(4 \cos \theta _1+\cos 2\theta_1+5\big) \cos \theta _2 \alpha _2 \chi ^2-6
	\cos \theta _2 \alpha _2+2 \cos \theta _1 \big(-3 \cos 2\theta_1 \alpha _1 \chi ^2\nonumber\\&+\big(2 \sqrt{6} \sin
	\theta _1 \alpha _1+3 \alpha _1-6 \alpha _3\big) \chi ^2+6 \alpha _1\big)+12 \alpha _3\big)\big)^2+D_2 \big(3 \big(\big(\big(3
	\chi ^2+2\big) \cos 2\theta_2+\cos 2\theta_1 \big(-\cos 2\theta_2 \chi ^2\nonumber\\&+3 \chi ^2+2\big)\big) \chi
	^2+\big(\chi ^2+2\big) \big(7 \chi ^2+6\big)\big) p_{\theta_1}+\big(\chi ^3+\chi \big) \big(2 \sin \theta _1 \big(-\big(\cos3
\theta _2-2 \cos 2\theta_2\big) \alpha _2 \chi ^2\nonumber\\&+2 \alpha _2 \chi ^2+4 \alpha _3+\cos \theta _2 \big(\big(\alpha
	_2+4 \alpha _3\big) \chi ^2+4 \alpha _2\big)\big)-\big(\big(\cos 2\theta_2-4 \cos \theta _2\big) \chi ^2+5 \chi
	^2+2\big) \sin 2\theta_1 \alpha _1\big)\big)^2\nonumber\\&-8 \cos \theta _1 \big(\chi ^2 \cos \theta _1-1\big) \cos
	\theta _2 \big(\cos \theta _2 \chi ^2+1\big) \alpha _1 \alpha _2-8 \cos \theta _1 \big(\chi ^2 \cos \theta
	_1-1\big) \big(\cos \theta _2 \chi ^2+1\big) \alpha _1 \alpha _3\nonumber\\&-8 \big(\chi ^2 \cos \theta _1-1\big) \cos
	\theta _2 \big(\cos \theta _2 \chi ^2+1\big) \alpha _2 \alpha _3\big) \chi ^2+\big(-\chi ^2 \big(\cos 2 \theta
	_2 \chi ^2+5 \chi ^2+6\big) \cos ^2\theta _1-4 \chi ^2 \sin ^2\theta _1\nonumber\\&-4 \chi ^2 \sin ^2\theta _2-\cos
	^2\theta _2 \big(4 \sin ^2\theta _1 \chi ^4+6 \chi ^2\big)-6\big) \big(-6 \big(\cos 2\theta_2 \chi ^2-3
	\chi ^2-2\big) \sin 2\theta_1 J_t^2 \chi ^2\nonumber\\&+4 \sin \theta _1 \big(\chi ^2 \cos ^3\theta _2+\cos ^2\theta
	_2-\cos \theta _1 \sin ^2\theta _2 \big(2 \chi ^2 \sin ^2\theta _2+1\big)\big) \alpha _2^2 \chi ^2+4
	\big(\cos \theta _2 \chi ^2\nonumber\\&+1\big) \sin \theta _1 \alpha _3^2 \chi ^2+2 \sin 2\theta_1 D_3 \big(9
	\big(\big(\big(3 \chi ^2+2\big) \cos 2\theta_2+\cos 2\theta_1 \big(-\cos 2\theta_2 \chi ^2+3 \chi
	^2+2\big)\big) \chi ^2\nonumber\\&+\big(\chi ^2+2\big) \big(7 \chi ^2+6\big)\big) p_{\theta_2}+2 \big(\chi ^3+\chi \big) \sin \theta _2 \big(-3
	\big(4 \cos \theta _1+\cos 2\theta_1+5\big) \cos \theta _2 \alpha _2 \chi ^2-6 \cos \theta _2
	\alpha _2\nonumber\\&+2 \cos \theta _1 \big(-3 \cos 2\theta_1 \alpha _1 \chi ^2+\big(2 \sqrt{6} \sin \theta _1 \alpha _1+3
	\alpha _1-6 \alpha _3\big) \chi ^2+6 \alpha _1\big)+12 \alpha _3\big)\big){}^2 \chi ^2\nonumber\\&+4 \cos \theta _2 \big(\cos \theta
	_2 \chi ^2+1\big) \sin 2\theta_1 \alpha _1 \alpha _2 \chi ^2+4 \big(\cos \theta _2 \chi ^2+1\big) \sin 2
	\theta _1 \alpha _1 \alpha _3 \chi ^2+8 \cos \theta _2 \big(\cos \theta _2 \chi ^2\nonumber\\&+1\big) \sin \theta _1
	\alpha _2 \alpha _3 \chi ^2+4 \cos \theta _1 \sin \theta _1 \big(4 \sin ^2\theta _1 \chi ^2+2 \sin ^2\theta
	_2 \chi ^2-2 \cos \theta _2 \chi ^2+\cos ^2\theta _2 \big(4 \chi ^2 \sin ^2\theta _1\nonumber\\&+3\big) \chi ^2+3 \cos
	\theta _1 \big(\cos \theta _2 \chi ^4+\chi ^2\big)+1\big) \alpha _1^2+8 \big(\chi ^2 \cos \theta _1-1\big)
	\cos \theta _2 \big(\cos \theta _2 \chi ^2+1\big) \sin \theta _1 \alpha _1 \alpha _2\nonumber\\&+8 \big(\chi ^2 \cos
	\theta _1-1\big) \big(\cos \theta _2 \chi ^2+1\big) \sin \theta _1 \alpha _1 \alpha _3+2 D_4 \big(18
	\big(\cos 2\theta_2 \chi ^2-3 \chi ^2-2\big) \sin 2\theta_1 p_{\theta_2} \chi ^2+2 \big(\chi ^3\nonumber\\&+\chi \big) \sin \big(\theta
	_2\big) \big(4 \cos ^2\theta _1 \big(6 \sin \theta _1+\sqrt{6}\big) \alpha _1 \chi ^2+6 \cos \theta _2 \big(2
	\sin \theta _1+\sin 2\theta_1\big) \alpha _2 \chi ^2\nonumber\\&-2 \sin \theta _1 \big(-3 \cos 2\theta_1
	\alpha _1 \chi ^2+\big(2 \sqrt{6} \sin \theta _1 \alpha _1+3 \alpha _1-6 \alpha _3\big) \chi ^2+6 \alpha _1\big)\big)\big) \big(9
	\big(\big(\big(3 \chi ^2+2\big) \cos 2\theta_2\nonumber\\&+\cos 2\theta_1 \big(-\cos 2\theta_2 \chi ^2+3 \chi
	^2+2\big)\big) \chi ^2+\big(\chi ^2+2\big) \big(7 \chi ^2+6\big)\big) p_{\theta_2}+2 \big(\chi ^3+\chi \big) \sin \theta _2 \big(-3
	\big(4 \cos \theta _1\nonumber\\&+\cos 2\theta_1+5\big) \cos \theta _2 \alpha _2 \chi ^2-6 \cos \theta _2
	\alpha _2+2 \cos \theta _1 \big(-3 \cos 2\theta_1 \alpha _1 \chi ^2+\big(2 \sqrt{6} \sin \theta _1 \alpha _1+3
	\alpha _1\nonumber\\&-6 \alpha _3\big) \chi ^2+6 \alpha _1\big)+12 \alpha _3\big)\big)+2 D_2 \big(6 \big(\cos 2\theta_2 \chi ^2-3 \chi
	^2-2\big) \sin 2\theta_1 p_{\theta_1} \chi ^2+\big(\chi ^3\nonumber\\&+\chi \big) \big(2 \cos \theta _1 \big(-\big(\cos \big(3 \theta
	_2\big)-2 \cos 2\theta_2\big) \alpha _2 \chi ^2+2 \alpha _2 \chi ^2+4 \alpha _3+\cos \theta _2 \big(\big(\alpha _2+4
	\alpha _3\big) \chi ^2+4 \alpha _2\big)\big)\nonumber\\&-2 \cos 2\theta_1 \big(\big(\cos 2\theta_2-4 \cos \big(\theta
	_2\big)\big) \chi ^2+5 \chi ^2+2\big) \alpha _1\big)\big) \big(3 \big(\big(\big(3 \chi ^2+2\big) \cos 2\theta_2+\cos 2
	\theta _1 \big(-\cos 2\theta_2 \chi ^2\nonumber\\&+3 \chi ^2+2\big)\big) \chi ^2+\big(\chi ^2+2\big) \big(7 \chi ^2+6\big)\big)
	p_{\theta_1}+\big(\chi ^3+\chi \big) \big(2 \sin \theta _1 \big(-\big(\cos3 \theta _2-2 \cos 2\theta_2\big) \alpha
	_2 \chi ^2\nonumber\\&+2 \alpha _2 \chi ^2+4 \alpha _3+\cos \theta _2 \big(\big(\alpha _2+4 \alpha _3\big) \chi ^2+4 \alpha
	_2\big)\big)-\big(\big(\cos 2\theta_2-4 \cos \theta _2\big) \chi ^2+5 \chi ^2+2\big) \sin 2\theta_1
	\alpha _1\big)\big)\big)\big).
\end{align}

\begin{align}
\mathcal{F}_2&=\big(\chi ^2+1\big) D_5 \bigg(\chi ^2 \big(\cos  2\theta _1 \chi ^2-3 \chi ^2-2\big) \sin  2\theta _2 \big(-6 \big(\chi ^2
\big(\cos  2\theta _2 \chi ^2+5 \chi ^2+6\big) \cos ^2 \theta _1\nonumber\\&+4 \chi ^2 \sin ^2 \theta _1+4 \chi ^2 \sin
^2 \theta _2+\cos ^2 \theta _2 \big(4 \sin ^2 \theta _1 \chi ^4+6 \chi ^2\big)+6\big) J_t^2+2 \big(-2 \big(\cos
 \theta _2 \chi ^4\nonumber\\&+\chi ^2\big) \cos ^3 \theta _1+2 \big(\cos  \theta _2 \chi ^2+1\big) \cos ^2\theta
_1+\sin ^2 \theta _1 \big(\chi ^2 \big(2 \chi ^2 \sin ^2 \theta _1+3\big) \cos ^2 \theta _2+2 \chi ^2 \sin
^2 \theta _1\nonumber\\&+2 \chi ^2 \sin ^2 \theta _2+3\big)\big) \alpha _1^2+2 \big(-2 \chi ^2 \big(\chi ^2 \cos \theta
_1-1\big) \cos ^3 \theta _2+\big(2-2 \chi ^2 \cos  \theta _1\big) \cos ^2 \theta _2\nonumber\\&+\sin ^2 \theta _2 \big(\chi ^2 \big(2 \chi ^2 \sin ^2 \theta _2+3\big) \cos ^2 \theta _1+2 \chi ^2 \sin ^2 \theta _1+2 \chi
^2 \sin ^2 \theta _2+3\big)\big) \alpha _2^2-4 \big(\chi ^2 \cos  \theta _1-1\big) \big(\cos  \theta _2 \chi
^2\nonumber\\&+1\big) \alpha _3^2+D_4 \big(9 \big(\big(\big(3 \chi ^2+2\big) \cos  2\theta _2+\cos  2\theta _1 \big(-\cos 2
\theta _2 \chi ^2+3 \chi ^2+2\big)\big) \chi ^2+\big(\chi ^2+2\big) \big(7 \chi ^2+6\big)\big) p_{\theta_2}\nonumber\\&+2 \big(\chi ^3+\chi \big) \sin
 \theta _2 \big(-3 \big(4 \cos  \theta _1+\cos  2\theta _1+5\big) \cos  \theta _2 \alpha _2 \chi ^2-6
\cos  \theta _2 \alpha _2+2 \cos  \theta _1 \big(-3 \cos  2\theta _1 \alpha _1 \chi ^2\nonumber\\&+\big(2 \sqrt{6} \sin
 \theta _1 \alpha _1+3 \alpha _1-6 \alpha _3\big) \chi ^2+6 \alpha _1\big)+12 \alpha _3\big)\big)^2+D_2 \big(3 \big(\big(\big(3
\chi ^2+2\big) \cos  2\theta _2+\cos  2\theta _1 \big(-\cos  2\theta _2 \chi ^2\nonumber\\&+3 \chi ^2+2\big)\big) \chi
^2+\big(\chi ^2+2\big) \big(7 \chi ^2+6\big)\big) p_{\theta_1}+\big(\chi ^3+\chi \big) \big(2 \sin  \theta _1 \big(-\big(\cos 3
\theta _2-2 \cos  2\theta _2\big) \alpha _2 \chi ^2+2 \alpha _2 \chi ^2\nonumber\\&+4 \alpha _3+\cos  \theta _2 \big(\big(\alpha
_2+4 \alpha _3\big) \chi ^2+4 \alpha _2\big)\big)-\big(\big(\cos  2\theta _2-4 \cos  \theta _2\big) \chi ^2+5 \chi
^2+2\big) \sin  2\theta _1 \alpha _1\big)\big)^2\nonumber\\&-8 \cos  \theta _1 \big(\chi ^2 \cos  \theta _1-1\big) \cos
 \theta _2 \big(\cos  \theta _2 \chi ^2+1\big) \alpha _1 \alpha _2-8 \cos  \theta _1 \big(\chi ^2 \cos \theta
_1-1\big) \big(\cos  \theta _2 \chi ^2+1\big) \alpha _1 \alpha _3\nonumber\\&-8 \big(\chi ^2 \cos  \theta _1-1\big) \cos
 \theta _2 \big(\cos  \theta _2 \chi ^2+1\big) \alpha _2 \alpha _3\big)-\big(\chi ^2 \big(\cos  2\theta _2 \chi
^2+5 \chi ^2+6\big) \cos ^2 \theta _1+4 \chi ^2 \sin ^2 \theta _1\nonumber\\&+4 \chi ^2 \sin ^2 \theta _2+\cos ^2\theta
_2 \big(4 \sin ^2 \theta _1 \chi ^4+6 \chi ^2\big)+6\big) \big(-6 \big(\cos  2\theta _1 \chi ^2-3 \chi ^2-2\big)
\sin  2\theta _2 J_t^2 \chi ^2\nonumber\\&+4 \big(\chi ^2 \cos ^3 \theta _1-\cos ^2 \theta _1-\cos  \theta _2 \sin
^2 \theta _1 \big(2 \chi ^2 \sin ^2 \theta _1+1\big)\big) \sin  \theta _2 \alpha _1^2 \chi ^2+4 \big(\chi ^2 \cos
 \theta _1-1\big) \sin  \theta _2 \alpha _3^2 \chi ^2\nonumber\\&+2 \sin  2\theta _2 D_6 \big(3 \big(\big(\big(3 \chi
^2+2\big) \cos  2\theta _2+\cos  2\theta _1 \big(-\cos  2\theta _2 \chi ^2+3 \chi ^2+2\big)\big) \chi
^2+\big(\chi ^2+2\big) \big(7 \chi ^2\nonumber\\&+6\big)\big) p_{\theta_1}+\big(\chi ^3+\chi \big) \big(2 \sin  \theta _1 \big(-\big(\cos 3
\theta _2-2 \cos  2\theta _2\big) \alpha _2 \chi ^2+2 \alpha _2 \chi ^2+4 \alpha _3+\cos  \theta _2 \big(\big(\alpha
_2+4 \alpha _3\big) \chi ^2\nonumber\\&+4 \alpha _2\big)\big)-\big(\big(\cos  2\theta _2-4 \cos  \theta _2\big) \chi ^2+5 \chi
^2+2\big) \sin  2\theta _1 \alpha _1\big)\big){}^2 \chi ^2+4 \cos  \theta _1 \big(\chi ^2 \cos \theta
_1\nonumber\\&-1\big) \sin  2\theta _2 \alpha _1 \alpha _2 \chi ^2+8 \cos  \theta _1 \big(\chi ^2 \cos \theta
_1-1\big) \sin  \theta _2 \alpha _1 \alpha _3 \chi ^2+4 \big(\chi ^2 \cos  \theta _1-1\big) \sin 2 \theta
_2 \alpha _2 \alpha _3 \chi ^2\nonumber\\&+4 \cos  \theta _2 \sin  \theta _2 \big(2 \sin ^2 \theta _1 \chi ^2+4 \sin
^2 \theta _2 \chi ^2-3 \cos  \theta _2 \chi ^2+\cos  \theta _1 \big(3 \cos  \theta _2 \chi ^2+2\big) \chi
^2+\cos ^2 \theta _1 \big(4 \chi ^2 \sin ^2 \theta _2\nonumber\\&+3\big) \chi ^2+1\big) \alpha _2^2+8 \cos  \theta _1
\big(\chi ^2 \cos  \theta _1-1\big) \big(\cos  \theta _2 \chi ^2+1\big) \sin  \theta _2 \alpha _1 \alpha _2+8
\big(\chi ^2 \cos  \theta _1\nonumber\\&-1\big) \big(\cos  \theta _2 \chi ^2+1\big) \sin  \theta _2 \alpha _2 \alpha _3+2 D_4
\big(18 \big(\cos  2\theta _1 \chi ^2-3 \chi ^2-2\big) \sin  2\theta _2 p_{\theta_2} \chi ^2+6 \big(\chi ^2+1\big) \big(4 \cos
 \theta _1 \chi ^2\nonumber\\&+\cos  2\theta _1 \chi ^2+5 \chi ^2+2\big) \sin ^2 \theta _2 \alpha _2 \chi +2 \big(\chi ^3+\chi
\big) \cos  \theta _2 \big(-3 \big(4 \cos  \theta _1+\cos  2\theta _1+5\big) \cos  \theta _2 \alpha
_2 \chi ^2\nonumber\\&-6 \cos  \theta _2 \alpha _2+2 \cos  \theta _1 \big(-3 \cos  2\theta _1 \alpha _1 \chi ^2+\big(2 \sqrt{6}
\sin  \theta _1 \alpha _1+3 \alpha _1-6 \alpha _3\big) \chi ^2+6 \alpha _1\big)\nonumber\\&+12 \alpha _3\big)\big) \big(9 \big(\big(\big(3 \chi
^2+2\big) \cos  2\theta _2+\cos  2\theta _1 \big(-\cos  2\theta _2 \chi ^2+3 \chi ^2+2\big)\big) \chi
^2+\big(\chi ^2+2\big) \big(7 \chi ^2+6\big)\big) p_{\theta_2}\nonumber\\&+2 \big(\chi ^3+\chi \big) \sin  \theta _2 \big(-3 \big(4 \cos \theta
_1+\cos  2\theta _1+5\big) \cos  \theta _2 \alpha _2 \chi ^2-6 \cos  \theta _2 \alpha _2+2 \cos \big(\theta
_1\big) \big(-3 \cos  2\theta _1 \alpha _1 \chi ^2\nonumber\\&+\big(2 \sqrt{6} \sin  \theta _1 \alpha _1+3 \alpha _1-6 \alpha _3\big)
\chi ^2+6 \alpha _1\big)+12 \alpha _3\big)\big)+2 D_2 \big(6 \big(\cos  2\theta _1 \chi ^2-3 \chi ^2-2\big) \sin 2 \theta
_2 p_{\theta_1} \chi ^2\nonumber\\&+\big(\chi ^3+\chi \big) \big(2 \sin  2\theta _1 \big(\sin  2\theta _2-2 \sin \theta
_2\big) \alpha _1 \chi ^2+4 \sin  \theta _1 \sin  \theta _2 \big(-4 \cos  \theta _2 \alpha _2 \chi ^2+3 \cos
 2\theta _2 \alpha _2 \chi ^2\nonumber\\&+\alpha _2 \chi ^2-2 \alpha _3 \chi ^2-2 \alpha _2\big)\big)\big) \big(3 \big(\big(\big(3 \chi
^2+2\big) \cos  2\theta _2+\cos  2\theta _1 \big(-\cos  2\theta _2 \chi ^2+3 \chi ^2+2\big)\big) \chi
^2+\big(\chi ^2+2\big) \big(7 \chi ^2\nonumber\\&+6\big)\big) p_{\theta_1}+\big(\chi ^3+\chi \big) \big(2 \sin  \theta _1 \big(-\big(\cos 3
\theta _2-2 \cos  2\theta _2\big) \alpha _2 \chi ^2+2 \alpha _2 \chi ^2+4 \alpha _3+\cos  \theta _2 \big(\big(\alpha
_2+4 \alpha _3\big) \chi ^2\nonumber\\&+4 \alpha _2\big)\big)-\big(\big(\cos  2\theta _2-4 \cos  \theta _2\big) \chi ^2+5 \chi
^2+2\big) \sin  2\theta _1 \alpha _1\big)\big)\big)\bigg).
\end{align}


\newpage

	\begin{thebibliography}{99}\label{bibliography}
		
		\bibitem{AdS/CFT chaos}
	L.~A.~Pando Zayas and C.~A.~Terrero-Escalante,
	Chaos in the Gauge / Gravity Correspondence,''
	JHEP \textbf{09} (2010), 094
	doi:10.1007/JHEP09(2010)094
	[arXiv:1007.0277 [hep-th]].
		
	\bibitem{maldacena}
		The Large N limit of superconformal field theories and supergravity,''
Adv. Theor. Math. Phys. \textbf{2} (1998), 231-252
doi:10.4310/ATMP.1998.v2.n2.a1
[arXiv:hep-th/9711200 [hep-th]].
		
		\bibitem{Witten}
		
	Anti-de Sitter space and holography,''
Adv. Theor. Math. Phys. \textbf{2} (1998), 253-291
doi:10.4310/ATMP.1998.v2.n2.a2
[arXiv:hep-th/9802150 [hep-th]].
		
		\bibitem{GKP}
		Gauge theory correlators from noncritical string theory,''
Phys. Lett. B \textbf{428} (1998), 105-114
doi:10.1016/S0370-2693(98)00377-3
[arXiv:hep-th/9802109 [hep-th]].
		
		\bibitem{PDA1}Analytic Non-integrability in String Theory,''
Phys. Rev. D \textbf{84} (2011), 046006
doi:10.1103/PhysRevD.84.046006
[arXiv:1105.2540 [hep-th]].
		
		\bibitem{PDA2}
		Chaos rules out integrability of strings on AdS$_5 \times T^{1,1}$,''
Phys. Lett. B \textbf{700} (2011), 243-248
doi:10.1016/j.physletb.2011.04.063
[arXiv:1103.4107 [hep-th]].
		
		\bibitem{g1}
		K.~S.~Rigatos,
		Nonintegrability of $L^{a,b,c}$ quiver gauge theories,
		Phys. Rev. D \textbf{102} (2020) no.10, 106022
		doi:10.1103/PhysRevD.102.106022
		[arXiv:2009.11878 [hep-th]]
		
		\bibitem{bena}
		Hidden symmetries of the $AdS_5 \times S^5 $superstring,''
Phys. Rev. D \textbf{69} (2004), 046002
doi:10.1103/PhysRevD.69.046002
[arXiv:hep-th/0305116 [hep-th]].



  \bibitem{coni1}
D.~N.~Page and C.~N.~Pope,
Which Compactifications of $D=11$ Supergravity Are Stable?,
Phys. Lett. B \textbf{144} (1984), 346-350
doi:10.1016/0370-2693(84)91275-9
		\bibitem{coni2}
		New Compactifications of Chiral $N=2 d=10$ Supergravity,''
Phys. Lett. B \textbf{153} (1985), 392-396
doi:10.1016/0370-2693(85)90479-4.
		
		
		
		\bibitem{coni3}
		P.~Candelas and X.~C.~de la Ossa,
Comments on Conifolds,
Nucl. Phys. B \textbf{342} (1990), 246-268
doi:10.1016/0550-3213(90)90577-Z.
		
			\bibitem{analitic1}
	J. J. Kovacic, “An algorithm for solving second order linear homogeneous differential equa-
	tions,” J.Symb.Comput. 2 (1986) 3
		
			\bibitem{analitic2}
	B. D. Saunders, “An implementation of Kovacic’s algorithm for solving second order linear
	homogeneous differential equations,” in: The Proceedings of the 4th ACM Symposium on
	Symbolic and Algebraic Computation, SYMSAC’81, August 5–7, Snowbird, USA, 1981
		
			\bibitem{analitic3}
	J. Kovacic, “Picard-Vessiot Theory, Algebraic Groups and Group Schemes,” Depart-
	ment of Mathematics, the City College of the City University of New York, 2005,
	https://ksda.ccny.cuny.edu/PostedPapers/pv093005.pdf
	
	
		\bibitem{hY1}
	I.~Kawaguchi, T.~Matsumoto and K.~Yoshida,
	Jordanian deformations of the $AdS_5 \times S^5$ superstring,
	JHEP \textbf{04} (2014), 153
	doi:10.1007/JHEP04(2014)153
	[arXiv:1401.4855 [hep-th]].
	
	
		\bibitem{hY2}
	T.~Matsumoto and K.~Yoshida,
	Yang\textendash{}Baxter sigma models based on the CYBE,
	Nucl. Phys. B \textbf{893} (2015), 287-304
	doi:10.1016/j.nuclphysb.2015.02.009
	[arXiv:1501.03665 [hep-th]].
	
	\bibitem{mY1}
	C.~Klimcik,
	Yang-Baxter sigma models and dS/AdS T duality,
	JHEP \textbf{12} (2002), 051
	doi:10.1088/1126-6708/2002/12/051
	[arXiv:hep-th/0210095 [hep-th]]
	
		\bibitem{mY2}
C.~Klimcik,
On integrability of the Yang-Baxter sigma-model,
J. Math. Phys. \textbf{50} (2009), 043508
doi:10.1063/1.3116242
[arXiv:0802.3518 [hep-th]].


\bibitem{scY1}
F.~Delduc, M.~Magro and B.~Vicedo,
On classical $q$-deformations of integrable sigma-models,
JHEP \textbf{11} (2013), 192
doi:10.1007/JHEP11(2013)192
[arXiv:1308.3581 [hep-th]].


\bibitem{scY2}
F.~Delduc, M.~Magro and B.~Vicedo,
An integrable deformation of the $AdS_5 \times S^5$ superstring action,
Phys. Rev. Lett. \textbf{112} (2014) no.5, 051601
doi:10.1103/PhysRevLett.112.051601
[arXiv:1309.5850 [hep-th]].

\bibitem{scY3}
F.~Delduc, M.~Magro and B.~Vicedo,
Derivation of the action and symmetries of the $q$-deformed $AdS_{5} \times S^{5}$ superstring,
JHEP \textbf{10} (2014), 132
doi:10.1007/JHEP10(2014)132
[arXiv:1406.6286 [hep-th]].

\bibitem{analytic1}
D.~Roychowdhury,
Analytic integrability for strings on $ \eta $ and $ \lambda $ deformed backgrounds,
JHEP \textbf{10} (2017), 056
doi:10.1007/JHEP10(2017)056
[arXiv:1707.07172 [hep-th]].

\bibitem{chaos1}
A.~Banerjee and A.~Bhattacharyya,
Probing analytical and numerical integrability: the curious case of $ (AdS_{5}\times S^{5})_\eta$,
JHEP \textbf{11} (2018), 124
doi:10.1007/JHEP11(2018)124
[arXiv:1806.10924 [hep-th]].





\bibitem{D1}
P.~M.~Crichigno, T.~Matsumoto and K.~Yoshida,
Deformations of $T^{1,1}$ as Yang-Baxter sigma models,
JHEP \textbf{12} (2014), 085
doi:10.1007/JHEP12(2014)085
[arXiv:1406.2249 [hep-th]].
	
	
\bibitem{D2}
	J.~i.~Sakamoto and K.~Yoshida,
Yang-Baxter deformations of $W_{2,4}\times T^{1,1}$ and the associated T-dual models,
Nucl. Phys. B \textbf{921} (2017), 805-828
doi:10.1016/j.nuclphysb.2017.06.017
[arXiv:1612.08615 [hep-th]].
	
	\bibitem{D3}
	L.~Rado, V.~O.~Rivelles and R.~S\'anchez,
Yang-Baxter deformations of the $AdS_5$ $\times$ $T^{1,1}$ superstring and their backgrounds,
JHEP \textbf{02} (2021), 126
doi:10.1007/JHEP02(2021)126
[arXiv:2010.14081 [hep-th]].
	
	
	\bibitem{D4}
	 O.~Lunin and J.~M.~Maldacena,
Deforming field theories with U(1) x U(1) global symmetry and their gravity duals,
JHEP \textbf{05} (2005), 033
doi:10.1088/1126-6708/2005/05/033
[arXiv:hep-th/0502086 [hep-th]].
	
	\bibitem{D5}
	A.~Catal-Ozer,
Lunin-Maldacena deformations with three parameters,
JHEP \textbf{02} (2006), 026
doi:10.1088/1126-6708/2006/02/026
[arXiv:hep-th/0512290 [hep-th]].
	
		\bibitem{pani}
	K. L. Panigrahi and M. Samal, Chaos in classical string dynamics in  $\gamma$-deformed  $AdS_5 \times T^{1,1}$, Phys. Lett. B 761, 475-481 (2016) doi:10.1016/j.physletb.2016.08.021
[arXiv:1605.05638 [hep-th]].

\bibitem{Arutyunov}
G.~Arutyunov, C.~Bassi and S.~Lacroix,
New integrable coset sigma models,
JHEP \textbf{03} (2021), 062
doi:10.1007/JHEP03(2021)062
[arXiv:2010.05573 [hep-th]].

\bibitem{C1}
T. Ishii, S. Kushiro and K. Yoshida, “Chaotic string dynamics in deformed T1,1,” JHEP 05,
158 (2021) doi:10.1007/JHEP05(2021)158 [arXiv:2103.12416 [hep-th]].


\bibitem{C2}
J.~Pal, A.~Mukherjee, A.~Lala and D.~Roychowdhury,
Analytic (non)integrability of Arutyunov-Bassi-Lacroix model,
Phys. Lett. B \textbf{820} (2021), 136496
doi:10.1016/j.physletb.2021.136496
[arXiv:2106.01237 [hep-th]].


\bibitem{eta}
L. Rado, V. O. Rivelles and R. Sanchez, “Bosonic $\eta$-deformations of non-integrable back-
grounds,” JHEP 03, 094 (2022) doi:10.1007/JHEP03(2022)094 [arXiv:2111.13169 [hep-th]].
	\end{thebibliography}
\end{document}